\documentclass[aps,prc,preprint,showpacs,showkeys]{revtex4}
\usepackage{graphicx}
\usepackage{hyperref}
\usepackage{latexsym}
\usepackage{bm}
\usepackage{epsf}

\newcommand{\bra}{\langle}
\newcommand{\ket}{\rangle}

\begin{document}
    
\bibliographystyle{prsty2}    

\title{The three-body problem with short-range forces: renormalized
equations and regulator-independent results}

\author{I.\ R.\ Afnan}
\email[E-mail:]{Iraj.Afnan@flinders.edu.au}
\affiliation{School of Chemistry, Physics and Earth Sciences\\
Flinders University, GPO Box 2100, Adelaide 5001, Australia}
\author {Daniel R. Phillips}
\email[E-mail:]{phillips@phy.ohiou.edu}
\affiliation{Department of Physics and Astronomy,\\ 
Ohio University, Athens, OH 45701}

\begin{abstract}
We discuss effective field theory treatments of the problem of three
particles interacting via short-range forces. One case of such a
system is neutron-deuteron scattering at low energies. We demonstrate
that in attractive channels the renormalization-group evolution of the
1+2 scattering amplitude may be complicated by the presence of
eigenvalues greater than unity in the kernel. We also show that these
eigenvalues can be removed from the kernel by one subtraction,
resulting in an equation which is renormalization-group invariant. A
unique solution for 1+2 scattering phase shifts is then obtained. We
give an explicit demonstration of our procedure for both the case of
three spinless bosons and the case of the doublet channel in nd
scattering.  After the contribution of the two-body effective range is
included in the effective field theory, it gives a good description of
the nd doublet phase shifts below deuteron breakup threshold.
\end{abstract}

\maketitle

\section{Introduction}\label{sec1}

For over forty years now the three-nucleon problem has, with
considerable success, been used as a testing ground for the
nucleon-nucleon interaction.  Usually the three-body equations used
are based on the Schr\"odinger equation or its implementation for
scattering in the form of the Faddeev~\cite{Fa61} equations with the
two-body interaction being one of finite range. However, discussions
of the case in which the range of the interaction is significantly
less than the wavelengths of interest, i.e.  $k R \ll 1$, also have a
long history. There has been renewed interest in this case with the
advent of Effective Field Theory (EFT) descriptions of few-nucleon
systems at low energy~\cite{vK99,Be00,Ph01}. In an EFT treatment of
the problem of two- and three-body scattering at energies such that $k
R \ll 1$ the two-body scattering problem requires renormalization
since the leading-order two-body potential is a three-dimensional
delta function.

But in low-energy $NN$ scattering $k$ and $R$ are not the only scales
in the problem.  The presence of a low-energy bound state in the $NN$
system---the deuteron---means that we must also account for the
deuteron binding momentum $\gamma=\sqrt{-m\epsilon_d} \sim 1/a$---with
$a \gg R$ the unnaturally-large $NN$ scattering length---when we do an
EFT analysis of this problem. In technical terms the presence of an
enhanced two-body scattering length---or equivalently a
near-zero-energy bound state---means that there is a non-trivial fixed
point in the renormalization-group evolution of this leading-order
potential. As long as this is accounted for, and a power counting
built around the scale hierarchy:
\begin{equation}
k \sim \gamma \ll 1/R,
\end{equation}
a systematic EFT can be established and renormalized using a variety
of regularization
schemes~\cite{We90,We91,vK97,vK98,Ka98A,Ka98B,Ge98A,Bi99}.  A similar
scale hierarchy (and hence a similar EFT) governs the low-energy
interactions of Helium-4 atoms~\cite{Bd99A,Bd99B,BH02}. It is also
relevant to the physics of Bose-Einstein condensates, if the external
magnetic field is adjusted such that the atoms (e.g. ${}^{85}$Rb) are
near a Feshbach resonance~\cite{BH01,Br02}.

As a first step in extending this EFT to heavier nuclei, the
three-nucleon system was considered, and the Faddeev equations for the
particular case of a zero-range interaction were solved. It was soon
discovered that the leading-order (LO) EFT equation for the quartet
(total angular momentum $3/2$) channel yielded a unique
solution~\cite{BvK97,Bd98}, while for the doublet (total angular
momentum $1/2$) channel the corresponding equation did not yield a
unique solution---at least in the absence of three-body
forces~\cite{Bd99A,Bd99B,Bd00}. This could be simply understood on the
grounds that in the quartet channel the effective interaction between
the neutron and the deuteron is repulsive as a result of the Pauli
principle, and this ultimately means that the neutron and deuteron do
not experience a zero-range interaction. In contrast, in the doublet
channel the effective neutron-deuteron interaction is attractive and
the full difficulties of the zero-range interaction manifest
themselves. These difficulties were first elucidated by Thomas, who
pointed out that---if two-body forces alone are employed---the nuclear
force must have a finite range if the binding energy of nuclei is to
be finite~\cite{Th35}.

The three-body scattering problem for zero-range interactions
considered in the seminal work of Bedaque and
collaborators~\cite{BvK97,Bd98,Bd99A,Bd99B,Bd00} was first considered
in research that antedates Faddeev's landmark 1961 paper: by
Skorniakov and Ter-Martirosian~\cite{STM57} and by
Danilov~\cite{Da61}. These authors found similar difficulties to
Bedaque {\it et al.}, and traced the non-uniqueness to the fact that
in the asymptotic region this three-body equation for scattering
reduces to a homogeneous equation whose solution can be added to the
solution of the inhomogeneous equation with an arbitrary weighting---a
point recently reiterated by Blankleider and
Gegelia~\cite{Ge00,BlG00,BG01}.

In their 1999 papers~\cite{Bd99A,Bd99B}, Bedaque {\it et al.}
introduced a three-body force into the leading-order three-body EFT
equation, so as to obtain a unique solution for 1+2 phase shifts.
They adjusted this force in order to reproduce the experimental 1+2
scattering length.  The energy dependence of the 1+2 phase shift was
then predicted~\cite{Bd99A,Bd99B,Bd00}. The introduction of this three-body
force is unexpected if naive dimensional analysis is used to estimate
the size of various effects in the EFT, but it is apparently necessary
if the equations are to yield sensible, unique predictions for
physical observables. This also accords with the 1995 paper of
Adhikhari, Frederico, and Goldman, who pointed out that the
divergences in the kernel of the Faddeev equations for a zero-range
interaction may necessitate the introduction of a piece of three-body
data so that these divergences can be renormalized
away~\cite{Ad95}. (But see Refs.~\cite{Ge00,BlG00,BG01} for a
conflicting view.)

In an attempt to get some insight into alternative ways to establish a
unique solution to the three-body scattering problem at leading order
in the effective field theory, we try to bridge the gap between the
Faddeev approach---in which the interaction has a finite range---and
the EFT formulation of this problem.  In Sec.~\ref{sec2}, we examine
the Amado model~\cite{Am63} for the case of three spinless
bosons. Here we look at scattering in which the interaction of an
incident particle on a composite system of the other two is considered
within the framework of the Lagrangian for the Lee model~\cite{Le54,
VAA61}. If three-body forces are neglected then the only difference
between this approach and those at LO in the EFT of
Refs.~\cite{We90,We91,vK97,vK98,Ka98A,Ka98B,Ge98A,Bi99} is that in the
Lee model Lagrangian one may introduce a form factor that plays the
role of a cut-off in the theory. In this way we can connect the LO EFT
equations (without a three-body force) to those found in the Amado
model, by taking the limit as the range of the interaction goes to
zero. The resulting equation has a non-compact kernel unless a cutoff
is imposed on the momentum integration. We then reproduce and
reiterate the results of Refs.~\cite{STM57,Da61,Bd99A,Bd99B}, demonstrating
that the low-energy solution of the equation changes radically as the
cutoff is varied. Using a renormalization-group analysis we trace this
unreasonable cutoff dependence to the presence of eigenvalues equal
to 1 in the kernel of the integral equation.

In Section~\ref{sec3} we use a subtraction originally developed by
Hammer and Mehen~\cite{HM00} to remove these eigenvalues. Our analysis
of Sec.~\ref{sec2} then allows us to demonstrate that the subtracted
three-body equation is renormalization-group-invariant. The
subtraction of Ref.~\cite{HM00} was employed at a specific energy, and
used experimental data from the three-body system to determine the
half-off-shell behaviour of the 1+2-amplitude. Here we go further, and
show that using low-energy two-body data plus just one piece of
experimental data for the three-body system---the 1+2 scattering
length---we can predict the low-energy three-body phase shifts. We do
this by first solving the subtracted integral equation for the
half-off-shell threshold 1+2-amplitude. We then use this result to
derive unique predictions for the full off-shell behaviour of the
1+2-threshold-amplitude, and thence for the 1+2-amplitude at any
energy.  The equations derived in this way are equivalent to those of
Bedaque {\it et al.}, but represent a reformulation of the problem in
which only physical, renormalized quantities appear. In consequence
the leading-order three-body force of Refs.~\cite{Bd99A,Bd99B} does
not appear in our equations. Our single subtraction
ultimately allows us to generate predictions for the energy-dependence
of the 1+2 phase shifts at leading order in the EFT without the
presence of an explicit three-body force.  The subtraction does,
though, require data from the three-body system (namely the 1+2
scattering length) before other three-body observables can be
predicted.

In Sec.~\ref{sec4} we apply the formalism of
Secs.~\ref{sec2}--\ref{sec3} to the---conceptually identical but
technically more complicated---case of the doublet channel in nd
scattering. Here, we compare the numerical solution to our
once-subtracted equation with phase-shift data.  In Sec.~\ref{sec5} we
consider higher-order corrections to the LO EFT and illustrate that
the results from the EFT are in good agreement with the nd data below
three-nucleon breakup threshold if the sub-leading (two-body) terms in
the EFT expansion are adjusted so as to reproduce the asymptotic
S-state normalization of deuterium. The resulting description of the
doublet phase shifts is very good up to the deuteron breakup
threshold.  Finally in Sec.~\ref{sec6} we present some concluding
remarks regarding the limitations of this method and discuss the
convergence and usefulness of the EFT.

\section{Three-boson scattering at low energy}\label{sec2}

Consider a system of three bosons at energies so low that the details
of their interaction are not probed.  Suppose, in addition, that two
of the bosons can form a bound state---``the dimer'', with binding
energy $-\epsilon_d$.  In this section we will compute the amplitude for
boson-dimer scattering in a low-energy effective theory.  This
problem has been studied for almost 50 years~\cite{STM57}, and has
recently been revisited in the context of EFT~\cite{Bd99A,Bd99B,Ge00,HM01,Bd02}.
In Sec.~\ref{sec-amado} we derive the Faddeev equations for this
system, in order to establish the notation used elsewhere in the
paper. We then show in Sec.~\ref{sec-RG} that this equation is not
renormalization-group (RG) invariant, i.e. changing the regularization
procedure used in the integral equation alters its physical
predictions significantly.  In particular, we will demonstrate that
this lack of RG invariance is due to the existence of eigenvalues equal
to one in the kernel of the integral equation.

\subsection{The Amado equations in the limit of zero-range interactions}

\label{sec-amado}

Consider a field theory of bosons $N$, in which the two-boson bound
state (``dimer'') $D$ is included as an explicit degree of freedom.
In this model the Lagrangian can be written as~\cite{Ka97,BG00}:
\begin{equation}
{\cal L} = N^\dag \left(i \partial_0 + \frac{\nabla^2}{2M}\right)N +
D^\dag\,\Delta\,D - g\left[D^\dag N N + D N^\dag N^\dag\right] \
.\label{eq:2.1}
\end{equation}
Here $\Delta$ is the bare inverse free propagator for the dimer. This
is basically the Lee model~\cite{Le54} for $D\leftrightarrow N N$.
Historically, in order to obtain a finite amplitude for boson-dimer
scattering, a regularization scheme has been invoked.  This can be
achieved either through the introduction of a cut-off in all momentum
integrals or by including a form factor in the interaction Lagrangian
\textit{i.e.} replacing $g\rightarrow g(p)$, with $p$ is the relative
momentum of the two bosons in $D\rightarrow NN$. The Amado
model~\cite{VAA61, Am63} entails the second choice for the
regularization. As a result the equation for boson-dimer scattering,
after partial-wave expansion, takes the form~\cite{Am63}\footnote{This
is Eq.~(21) of Amado~\cite{Am63} written in the notation commonly
used, and is the Faddeev equation for a rank-one separable two-body
potential.}
\begin{equation}
X_\ell(q,q';E) = 2 Z_\ell(q,q';E) + \int \limits_0^\infty\,dq''\, q''^2\
2 Z_\ell(q,q'',E)\,\tau \left(E^{(+)}-\frac{q''^2}{2\nu}\right)\, X_\ell(q'',q';E)\ \ .
\label{eq:2.2}
\end{equation}
with $E^{(+)}=E + i \eta$, $\eta$ a positive infinitesimal.  The Born
term $Z_\ell(q,q';E)$ is the amplitude for one-boson exchange, and is given
by
\begin{equation}
Z_\ell(q,q';E) = \frac{\lambda}{2}\,\int\limits_{-1}^{+1}
dx\frac{g(K)\,g(Q)}{E -
\frac{1}{m}(q^2+q'^2+\vec{q}\cdot\vec{q\,}')}P_\ell(x)\ ,
\label{eq:2.3}
\end{equation}
where $x= \hat{q}\cdot\hat{q}'$, $P_\ell$ is the Legendre function of
order $\ell$ and $\lambda=1$ for three identical bosons. In
Eq.~(\ref{eq:2.3}) the relative momenta of the pair in the vertices
$D\leftrightarrow NN$ are given by
\begin{equation}
K = \left|\vec{q\,}'+ \frac{1}{2}\vec{q}\,\right|\quad\mbox{and}\quad
Q= \left|\vec{q}+\frac{1}{2}\vec{q\,}'\right|     \ .            \label{eq:2.4}
\end{equation}

Note that the convention of Lovelace~\cite{Lo64} for the recoupling
coefficient $\lambda$ differs from this by a factor of -1.  In fact,
as originally shown by Lovelace~\cite{Lo64}, Eqs.~(\ref{eq:2.2}) and
(\ref{eq:2.3}) also govern nd scattering in the $I=\frac{1}{2}$;
$S=\frac{3}{2}$ channel, i.e. the quartet, but with $\lambda=-1/2$ (in
the convention used in this work). The details of the recoupling
algebra for the bosonic, nd quartet, and nd $S=\frac{1}{2}$ channel,
are given in Appendix~\ref{app-recoupling}.

The off-shell two-body $NN$ amplitude for this Lagrangian is of the form
\begin{equation}
t(p,p';E) = g(p)\, \tau(E)\,g(p')\ ,                     \label{eq:2.5}
\end{equation}
where the dressed dimer propagator is given by
\begin{equation}
\tau(E) = \frac{S(E)}{E-\epsilon_d}\ , \label{eq:2.6}
\end{equation}
with $\epsilon_d$ the energy of the dimer bound-state
($\epsilon_d<0$). The function $S(E)$ is:
\begin{equation}
S(E) = \left[
\int\limits_0^\infty\,\frac{dp\,p^2\,g^2(p)}{(E-\frac{p^2}{m})(\epsilon_d-\frac{p^2}{m})}\right]^{-1}
\ ,  \label{eq:2.7}
\end{equation}
and the residue of the $NN$ propagator at the dimer pole is then clearly
$S(\epsilon_d)$.

At this stage the regularization function $g(p)$ is present in the
dressing of the dimer propagator $\tau(E)$, and also in the
one-nucleon exchange amplitude $Z_\ell(q,q';E)$. In writing
Eqs.~(\ref{eq:2.6}) and (\ref{eq:2.7}), we have imposed the
renormalization condition that the binding energy of the dimer takes
its physical value. After this renormalization is performed the
subtracted integral in Eq.~(\ref{eq:2.7}) is finite and the cut-off
function $g(p)$ can be taken to be one. In that limit
\begin{equation}
\tau(E)=\frac{S(E)}{E-\epsilon_d} = \frac{2}{\pi m^2}\ 
\frac{\gamma+\sqrt{-mE}}{E-\epsilon_d},
                            \ .                            \label{eq:2.8}
\end{equation}
with $\gamma \equiv \sqrt{-m \epsilon_d}$. The form of the two-body
amplitude (\ref{eq:2.5}) is---in the limit $g \rightarrow 1$---exactly
that obtained at leading order in an effective field theory with
short-range interactions
alone~\cite{We90,We91,vK97,vK98,Ka98A,Ka98B,Ge98A,Bi99}.

This allows us to write the integral equation for 1+2 scattering in
the limit $g(p)\rightarrow 1$ as
\begin{equation}
X_\ell(q,q';E) = 2 Z_\ell(q,q';E) + \int\limits_0^\infty dq''\,q''^2\,
2 Z_\ell(q,q'';E)\,
\frac{S\left(E^{(+)}-\frac{3q''^2}{4m}\right)}{E^{(+)}-\frac{3q''^2}{4m}-\epsilon_d}\,X_\ell(q'',q';E)
                                                            \ ,\label{eq:2.9}
\end{equation}
where, for $\ell=0$, the Born term $2 Z_0 (q,q';E)$ is given by
\begin{equation}
2 Z_0(q,q';E) \equiv {\cal Z}(q,q';E)
= -\lambda\,\frac{m}{qq'}\ln\left[\frac{ q^2 +q'^2+qq'-mE}{q^2+q'^2-
qq'-mE}\right]
\ ,                                                            \label{eq:2.10}
\end{equation}
and
\begin{equation}
S\left(E-\frac{3q^2}{4m}\right) \equiv S(E;q)
= \frac{2}{\pi m^2}\ \left[\,\gamma + \sqrt{\frac{3}{4}q^2
-mE}\,\right]                         \ . \label{eq:2.11}
\end{equation}
Here, we have included the factor of 2 resulting from the
symmetrization for identical particles in Eq.~(\ref{eq:2.2}) in the
definition of ${\cal Z}$.  These equations are identical to those employed in
the EFT of Refs.~\cite{Bd99B,Bd02}, although with different
normalization for the amplitude $X_\ell$. The relationship between our
conventions and those of Refs.~\cite{Bd99B,Bd02} is elucidated further
in Appendix~\ref{app-connect}.

In the above analysis we considered the integral equation for the
scattering amplitude or $T$-matrix. At any finite $k$ the kernel of
the integral equation has a pole, coming from the dimer propagator
(see Eq.~(\ref{eq:2.9})). Furthermore, above the breakup threshold,
the Born amplitude ${\cal Z}$ develops moving logarithmic singularities that
need to be dealt with when this equation is solved numerically, for
instance by performing a contour rotation that avoids these
singularities. In this work we restrict the analysis to energies below
the dimer-breakup threshold, and so the logarithmic $NNN$ cut is not
an issue. In this energy domain a reformulation of Eq.~(\ref{eq:2.9})
that eliminates the $D$ pole is useful.

The reformulation involves writing the $DN$ propagator:
\begin{equation}
\frac{1}{E^{(+)}-\epsilon_d-\frac{3q''^2}{4m}}={\cal P}\frac{1}{E - \epsilon_d
- \frac{3 q''^2}{4m}} - i \pi \delta \left(E - \epsilon_d - \frac{3
q''^2}{4m}\right)
\label{eq:2.12}
\end{equation}
We can then calculate the amplitude $X_0$ of Eq.~(\ref{eq:2.9}) by first
calculating the boson-dimer S-wave $K$-matrix using the integral
equation:
\begin{equation}
K(q,q';E) = {\cal Z}(q,q';E)+{\cal P} \int \limits_0^\infty dq''\, q''^2 \,
{\cal Z}(q,q'';E)\, \frac{S(E;q'')}{E - \epsilon_d -
\frac{3q''^2}{4m}}\,K(q'',q';E)\ .\label{eq:2.17}
\end{equation}
Below the three-body breakup threshold $K$ is a real symmetric matrix,
and this equation is free of singularities. It is therefore
numerically advantageous to solve this equation rather than
Eq.~(\ref{eq:2.9}), and this is the approach we have used in
generating our numerical results.

The relationship of the phase shifts to the on-shell K-matrix
$K(k,k;E)$, is provided by first employing the relation:
\begin{equation}
X_0(q,q';E)=K(q,q';E) - \frac{2 i m \pi k}{3} K(q,k;E) S(E;k) X_0(k,q';E).
\label{eq:XKreln}
\end{equation}
Then, to determine the S-wave on-shell scattering amplitude and
therefore the S-wave phase shift, we need to multiply the result from
Eq.~(\ref{eq:XKreln}) by the residue of the dimer propagator,
\textit{i.e.} define
\begin{equation}
T_0(q,q';E) = S^{1/2}(q,E)\,X_0(q,q';E)\,S^{1/2}(q',E) \ .  \label{eq:2.14}
\end{equation}
The boson-dimer S-wave scattering phase shifts are then related to the
amplitude $T_0$ at the on-shell point by:
\begin{equation}
T_0(k,k;E) = -\frac{3}{2m\pi k}\,e^{i\delta}\sin\delta  \ ,   \label{eq:2.15}
\end{equation}
where the on-shell momentum $k$ is defined by the relation
\begin{equation}
E-\frac{3k^2}{4m} = \epsilon_d.
\label{eq:2.16}
\end{equation} 
Using the relationships (\ref{eq:XKreln}) and (\ref{eq:2.15}) we find
that:
\begin{equation}
K(k,k;E)=-\frac{3m}{8 \gamma k} \tan \delta.
\label{eq:2.18}
\end{equation}

Since we will be using the boson-dimer
scattering length to renormalize our integral equation, we are particularly
interested in the case $E=\epsilon_d$ in Eq.~(\ref{eq:2.18}). Using
the boson-dimer effective-range expansion:
\begin{equation}
k\cot\delta = -\frac{1}{a_3} + \frac{1}{2} r_3 k^2+\cdots,
\label{eq:ere}
\end{equation}
with $a_3$ and $r_3$ being respectively, the boson-dimer scattering
length and effective range, we find:
\begin{equation}
K(0,0;\epsilon_d)=\frac{3m a_3}{8 \gamma}.
\label{eq:thold}
\end{equation}

\subsection{Renormalization-group invariance}

\label{sec-RG}

Now consider the convergence properties of the integrals in the
equation (\ref{eq:2.17}). If $q$ and $q''$ are both large
then ${\cal Z}$ behaves as $1/{q q''}$, while $S$ scales as
$q''$. Therefore the convergence of the integral in (\ref{eq:2.17})
depends on the behaviour of $X(q'',k;E)$ at large $q''$.  Perturbation
theory suggests that at large $q''$
\begin{equation}
K(q'',k;E) \sim {\cal Z}(q'',k;E) \sim 1/q''^2,
\end{equation}
and thus the integral equation will be well-behaved without
the need to impose any sort of regulator on the integral. However, this
conclusion is erroneous. 

In fact, the kernel of (\ref{eq:2.17}) has infinitely many eigenvalues
of order unity, as pointed out in Ref.~\cite{Da61}, and discussed in
detail by Amado and Noble~\cite{AN72}. The argument of Amado and Noble
may be heuristically paraphrased as follows. Calculating the trace of
the kernel of Eq.~(\ref{eq:2.17}) at $E=\epsilon_d$ we obtain:
\begin{equation}
{\rm tr(kernel)}=\frac{8 \lambda}{3 \pi} \int \limits_0^\infty
\frac{dq''}{q''} \ln \left(\frac{3q''^2 + \gamma^2}{q''^2 + \gamma^2}
\right) \left(\frac{\gamma}{q''} +
\sqrt{\frac{3}{4} + \left(\frac{\gamma}{q''}\right)^2}\right),
\end{equation}
an integral which diverges, and does so logarithmically. Since the
corrections proportional to $\left(\frac{\gamma}{q''}\right)^2$ do not
affect the ultraviolet behaviour of this integral, if a cutoff
$\Lambda$ is imposed we have
\begin{equation}
\rm{tr(kernel)} \rightarrow \frac{4 \ln 3}{\sqrt{3} \pi}
\ln (\Lambda)
\end{equation}
as $\Lambda \rightarrow \infty$. Simple power-counting arguments
demonstrate that the trace of all (positive-integer) powers of the
kernel contain a logarithmic divergence too. Since it is also the
case that the largest eigenvalue of the kernel is finite the only way
we can have
\begin{equation}
\rm{tr(kernel^n)} \sim \ln\left(\frac{\Lambda}{\gamma}\right)
\end{equation}
for all $n \geq 1$ is if the number of eigenvalues larger than one
grows logarithmically with $\Lambda$---a result in accord
with the analysis of Danilov~\cite{Da61} and Efimov~\cite{Ef70,Ef71}.

As we will discuss in more detail below, the presence of these
order-one eigenvalues means that the perturbation theory estimate of
$K$'s large-momentum behaviour is incorrect. Consequently a cutoff
must be imposed on the integral in (\ref{eq:2.17}), as otherwise the
integrals diverge. Once this is done the integral equation yields a
unique solution for a fixed value of $\Lambda$. In this work we impose
a sharp cutoff $\Lambda$, but choosing other, smoother cutoff
functions does not alter the essence of the following argument.

This regularization results in an equation:
\begin{equation}
K_\Lambda(q,q';E) = {\cal Z}(q,q';E)+{\cal P} \int \limits_0^\Lambda dq''\,
q''^2 \, {\cal Z}(q,q'';E)\, \frac{S(E - \frac{3 q''^2}{4m})}{E -
\epsilon_d - \frac{3q''^2}{4m}}\,K_\Lambda(q'',q';E)\ ,\label{eq:2.19}
\end{equation}
where the presence of the cutoff has made $K$ implicitly dependent on
$\Lambda$.  If the predictions of the theory are to be sensible we
must have
\begin{equation}
\Lambda \frac{d K_\Lambda(q,q';E)}{d\Lambda} \sim 0 \quad \mbox{for}
\quad q,q' \ll \Lambda.
\label{eq:2.20}
\end{equation}
In other words, the low-energy predictions of our boson-dimer
scattering calculation should not be affected by the imposition
of a cutoff at a momentum scale far above the ones that are physically
of interest. The demand (\ref{eq:2.20}) represents the renormalization group
(RG) for this problem.

Applying $\Lambda \frac{d}{d\Lambda}$ to both sides of Eq.~(\ref{eq:2.19})
we find that:
\begin{eqnarray}
\Lambda\frac{dK_\Lambda}{d\Lambda}(q,q';E) &=&  \Lambda^3
{\cal Z}(q,\Lambda;E) \frac{S(E;\Lambda)}{E-\epsilon_d- \frac{3\Lambda^2}{4m}} K_\Lambda(\Lambda,q';E)\nonumber \\ 
&+& {\cal P}\int\limits_0^\Lambda dq''\, q''^2 \, {\cal Z}(q,q'';E)\, \frac{S(E;q'')}{E -
\epsilon_d - \frac{3q''^2}{4m}}\, \Lambda
\frac{dK_\Lambda}{d\Lambda}(q'',q';E).
\label{eq:2.21}
\end{eqnarray}
For $q,k \ll \Lambda$ 
\begin{equation}
{\cal Z}(q,\Lambda;E) \sim \frac{1}{\Lambda^2}; \qquad S(E;\Lambda) \sim \Lambda,
\end{equation}
and, using the fact (which can be justified {\it a postieri}) that
$K(\Lambda,q';E) \sim \frac{1}{\Lambda}$ for $\Lambda \gg q'$ the first
term on the right hand side drops out of the integral equation (\ref{eq:2.21}) leaving:
\begin{equation}
\Lambda \frac{dK_\Lambda}{d\Lambda}(q,k;E) ={\cal
P}\int\limits_0^\Lambda dq''\, q''^2 {\cal Z}(q,q'';E)\,
\frac{S(E;q'')}{E-\epsilon_d-\frac{3q''^2}{4m}}\, \Lambda
\frac{dK_\Lambda}{d\Lambda}(q'',k;E).
\label{eq:2.22}
\end{equation}

At this point it is tempting to argue that since Eq.~(\ref{eq:2.22})
has a trivial solution, it follows that $K_\Lambda$ is RG invariant
(up to terms of $O(1/\Lambda)$). Such a conclusion is hasty, however.
Non-trivial solutions to Eq.~(\ref{eq:2.22}) exist if the kernel has
an eigenvalue of one. In fact, the presence of eigenfunctions of this
kernel corresponding to eigenvalue one has been proven rigorously in
the asymptotic regime $q \gg \gamma,k$ by Danilov~\cite{Da61} and by
Amado and Noble~\cite{AN72}. Here we repeat the analysis of Bedaque et
al.~\cite{Bd99A,Bd99B}, which demonstrates the presence of such an
eigenvector. The impact of this ``zero-mode'' on the spectrum of the
kernel has also been discussed by Gegelia and
Blankleider~\cite{BlG00,BG01} and by Bedaque et al.~\cite{Bd02}.

Consider $E=\epsilon_d$, i.e. scattering at the threshold for the $ND$
channel. Further, consider off-shell momenta $q$ such that $\gamma \ll
q \ll \Lambda$. In this limit we may neglect terms proportional to
$\gamma$ in $S(E;q)$ and ${\cal Z}(q,q'';E)$, and so the integrand 
becomes scale invariant:
\begin{equation}
\Lambda\frac{dK_\Lambda}{d\Lambda}(q,0;\epsilon_d) =\frac{4 \lambda}{\sqrt{3}
\pi} \frac{1}{q} \int\limits_0^\infty dq''\, \ln\left[\frac{q^2 + q''^2 + q
q''}{q^2 + q''^2 - q q''}\right] \Lambda
\frac{dK_\Lambda}{d\Lambda}(q'',0;\epsilon_d).
\label{eq:2.23}
\end{equation}
Inspired by the absence of any physical scale under the integral we
seek a power-law solution:
\begin{equation}
\Lambda \frac{d K_\Lambda}{d \Lambda}(q,0;\epsilon_d) \sim q^{s-1}
\end{equation}
As shown in Refs.~\cite{Da61,Bd99A,Bd99B}, such a solution exists provided that
$s$ obeys the transcendental equation:
\begin{equation}
\frac{8 \lambda}{\sqrt{3}s} \frac{\sin\left(\frac{\pi s}{6}\right)}
{\cos\left(\frac{\pi s}{2}\right)}=1.
\label{eq:Danilov}
\end{equation}
If $\lambda=1$ this equation has complex roots $s=\pm i s_0$ with
$s_0=1.0062$. Thus in this case the RG equation has non-trivial
solutions, which, provided $\Lambda \gg q \gg \gamma$, are of the form:
\begin{equation}
f(q)=\frac{1}{q} C \cos\left(s_0 \ln\left(\frac{q}{\Lambda}\right) 
+ \delta\right).
\label{eq:asf}
\end{equation}
Note that Eq.~(\ref{eq:2.23}) also governs $K(q,0;\epsilon_d)$,
at least for $\Lambda \gg q \gg \gamma$, and so a similar 
analysis applied to $K$ itself justifies the scaling 
$K(q,0;\epsilon_d) \sim 1/q$~\cite{Bd99A,Bd99B}.

\begin{figure}[h,b,t,p]
\centering\includegraphics[scale=0.6]{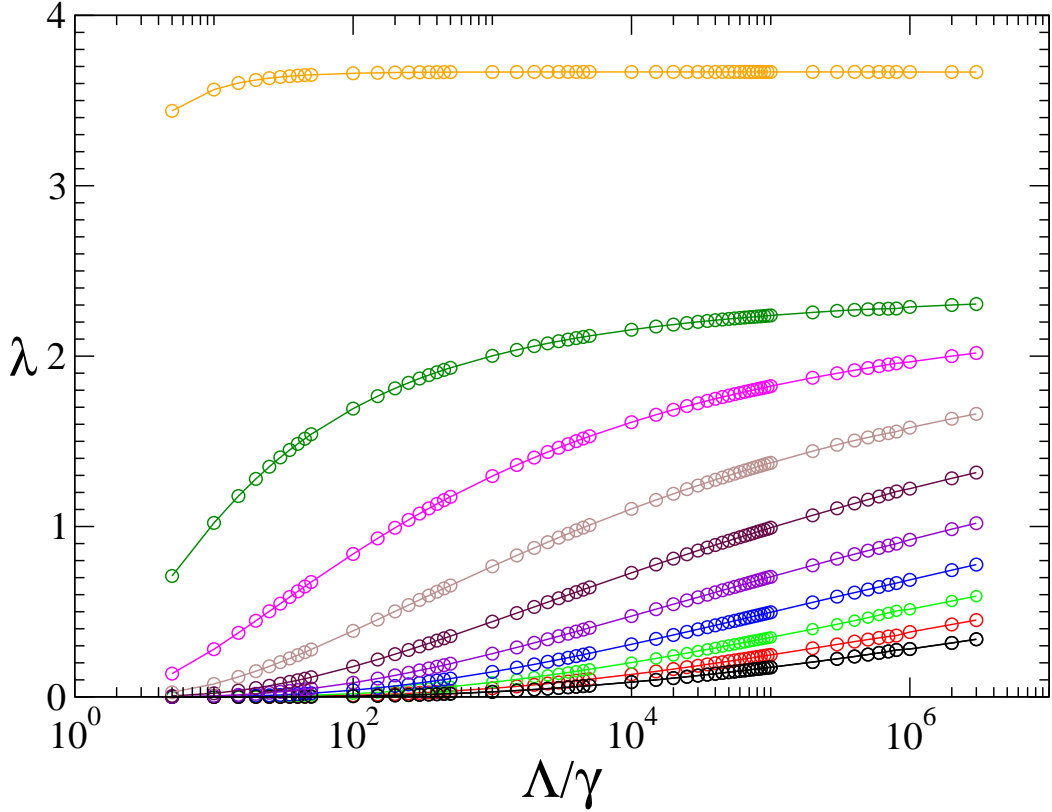}
\caption{The largest ten eigenvalues of the kernel of
Eq.~(\ref{eq:2.22}), for the case $\lambda=1$, and a range of
cutoffs.}
\label{fig-unsubtevals}
\end{figure}

For finite $\Lambda$ we can examine the eigenvalue spectrum of the
common kernel of Eqs.~(\ref{eq:2.19}) and (\ref{eq:2.22}).  Doing this
at $E=\epsilon_d \Leftrightarrow k=0$ and $\lambda=1$ for a variety of
cutoffs yields the results shown in Fig.~\ref{fig-unsubtevals}.  As
the cutoff increases there are more and more eigenvalues larger than
one, with a new eigenvalue of one appearing each time the cutoff
is increased by a factor of $e^{\pi/s_0}=22.7$. This corresponds 
to an increasing number of bound states of the
boson-dimer system described by
Eq.~(\ref{eq:2.19}), with the number of bound states growing
as~\cite{AN72}:
\begin{equation}
N=\frac{1.0062}{\pi}\ln\left(\frac{\Lambda}{\gamma}\right).
\end{equation}
This accumulation of zero-energy bound states in a system with
zero-range interactions was first pointed out by
Efimov~\cite{Ef70,Ef71} (see also Ref.~\cite{AN72}).

\begin{figure}[h,b,t,p]
\centering\includegraphics[scale=0.8]{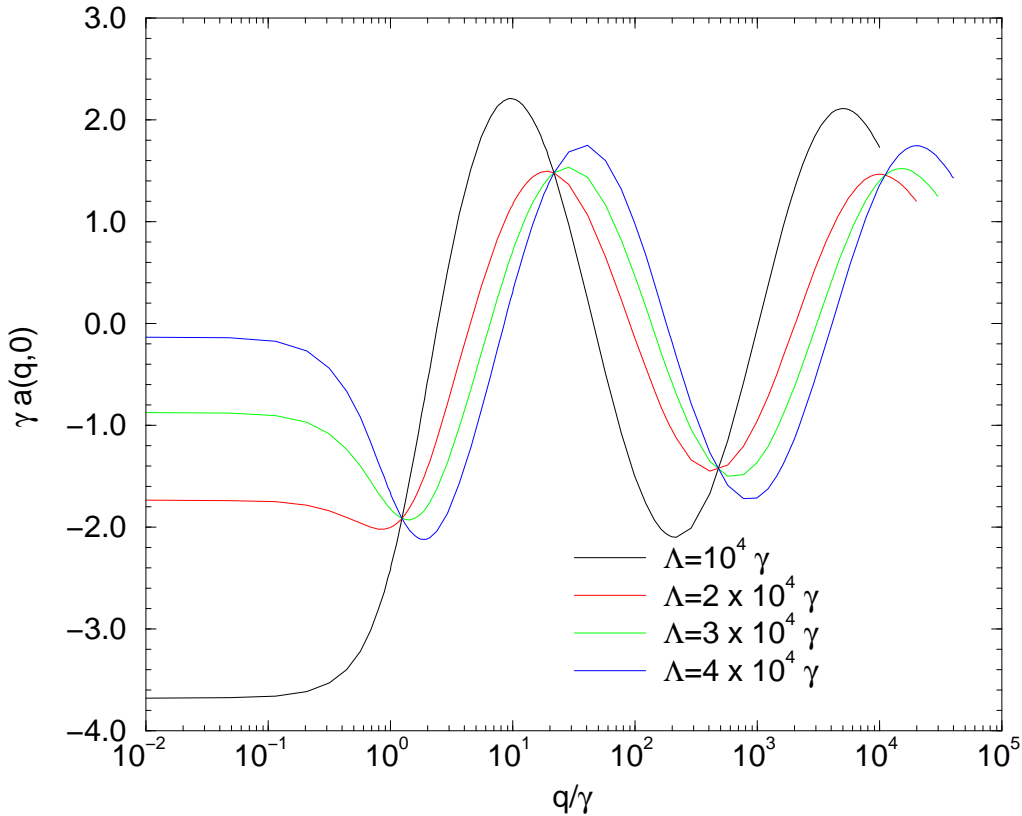}
\caption{Results for $a(q,0)$, as defined in Eq.~(\ref{eq:aeq}) when
various different cutoffs are imposed on the integral equation
(\ref{eq:2.17}).}
\label{fig-Kzeroen}
\end{figure}

The presence of these eigenvalues which cross one as the cutoff is
increased manifests itself as non-trivial cutoff dependence when
Eq.~(\ref{eq:2.19}) is solved. Some results found by solving this
equation for the half-off-shell amplitude (again at zero energy) are
displayed in Fig.~\ref{fig-Kzeroen}. Here $K(q,0;\epsilon_d) \sim
1/q$, and so we have chosen to present results for the quantity:
\begin{equation}
a(q,k) \equiv -\frac{2 \pi m S(E;q)}{3} X_0(q,k;E).
\label{eq:aeq}
\end{equation}
This also aids comparison with the results of Ref.~\cite{Bd99B}, with
which we are in complete agreement.  The renormalization-group
argument of this section ties the large changes in the low-momentum
amplitude seen in Fig.~\ref{fig-Kzeroen} to the spectrum of the kernel
of Eq.~(\ref{eq:2.19}), via the concomitant strong-RG evolution of
$K_\Lambda$ at low momentum.

Note that in contrast to the case $\lambda=1$, if
$\lambda=-\frac{1}{2}$ then the kernel of Eq.~(\ref{eq:2.19}) has no
eigenvalues larger than one, and so the renormalization-group
evolution of the K-matrix is smooth in that case. This means that as
$\Lambda \rightarrow \infty$ there is no cutoff-dependence in the
predictions for phase shifts---as was seen to be the case numerically
Ref.~\cite{BvK97,Bd98} where calculations of nd scattering in the quartet
channel were performed in the EFT. (See also Ref.~\cite{Bd02}.)  The
coupling at which the large-$\Lambda$ RG evolution becomes non-trivial
is the value at which Eq.~(\ref{eq:Danilov}) first develops
complex roots. This is~\cite{Da61}:
\begin{equation}
\lambda_c=\frac{3 \sqrt{3}}{4 \pi}.
\end{equation}

\section{The method of subtraction}\label{sec3}

Thus, as it stands, if $\lambda > \lambda_c$, Eq.~(\ref{eq:2.17})
needs additional renormalization before it can yield RG-invariant
predictions.  The solution proposed in Ref.~\cite{Bd99A,Bd99B} was to add a
counterterm to cancel the cutoff-dependence observed in
Fig.~\ref{fig-Kzeroen}.  The three-body force introduced to
renormalize the integral equation is not naively of the same order as
the terms in the EFT Lagrangian (\ref{eq:2.1}), but the analysis of
Ref.~\cite{Bd99A,Bd99B}, which has been recast in the previous section,
shows that it is necessary for renormalization. The naive dimensional
analysis estimate of the size of three-body forces is trumped by the
presence of the shallow bound state in the two-body system, which is
ultimately what leads to the Efimov spectrum shown in
Fig.~\ref{fig-unsubtevals}. Of course, as with any counterterm which
removes cutoff-dependence in a quantum field theory, a piece of
data is required to fix the value of the counterterm at a particular
scale. In Ref.~\cite{Bd99A,Bd99B} the boson-dimer scattering length $a_3$,
was chosen for this purpose.

More recently Blankleider and Gegelia~\cite{BlG00,BG01} have avoided
introducing a three-body force in the leading-order three-body EFT
equation by examining the solution of the homogeneous equation and
subtracting the oscillatory behavior.  However, in their work no
predictions for phase shifts were actually made. A subtraction
technique for the three-body problem with zero-range forces was also
suggested by Adhikari, Frederico, and Goldman~\cite{Ad95}. This
technique was implemented in another integral equation with a
non-compact kernel, that describing unregulated one-pion-exchange
between two nucleons~\cite{FTT99}. However, the subtraction in
Ref.~\cite{FTT99} is performed at large negative energy, and involves
demanding equivalence of the full and Born amplitudes at these
energies.

In this work we suggest an approach which is equivalent to that used
in Ref.~\cite{Bd99A,Bd99B}, but is formulated in an alternative fashion. Our
procedure involves a subtraction of the on-shell amplitude at
some---arbitrarily chosen but low---energy. The subtracted equation
has a unique solution, which is, up to corrections 
suppressed by $(p/\Lambda)^2$, RG-invariant. 

\subsection{Subtraction at threshold}\label{sec3.1}

Let us first consider the subtraction method applied to the integral
equation for the half-off-shell $1+2$ three boson threshold amplitude.  We use
information on the boson-dimer scattering length to fix the on-shell amplitude
at the threshold.  

Consider Eq.~(\ref{eq:2.17}) for the half off-shell amplitude at
$E=\epsilon_d$, \textit{i.e.}
\begin{equation}
K(q,0;\epsilon_d) = {\cal Z}(q,0;\epsilon_d) - \frac{4m}{3}
\int\limits_0^\Lambda dq''\,
{\cal Z}(q,q'';\epsilon_d)\,S(\epsilon_d;q'')\,K(q'',0;\epsilon_d) \ .
                      \label{eq:3.1}
\end{equation}
On the other hand, the on-shell amplitude at threshold should obey:
\begin{equation}
K(0,0;\epsilon_d) = \frac{3 m a_3}{8 \gamma} = {\cal Z}(0,0;\epsilon_d) -
\frac{4m}{3}\int\limits_0^\Lambda dq''\,{\cal Z}(0,q'';\epsilon_d)\ 
S(\epsilon_d; q'')\ K(q'',0;\epsilon_d)
                                   \ .\label{eq:3.2}
\end{equation}
If we now subtract Eq.~(\ref{eq:3.2}) from Eq.~(\ref{eq:3.1}), we get
an integral equation for the half-off-shell amplitude for which the
input is the boson-dimer scattering length $a_3$, in addition to the
two-body data, which in lowest order is just the binding energy of the
dimer. This equation is
\begin{equation}
K(q,0;\epsilon_d) = \frac{3 m a_3}{8 \gamma} + \Delta
{\cal Z}(q,0;\epsilon_d) - \frac{4m}{3}\int\limits_0^\Lambda dq''\, \Delta
{\cal Z}(q,q'';\epsilon_d)\,S(q'';\epsilon_d)\,K(q'',0;\epsilon_d),\
\label{eq:3.3}
\end{equation}
where
\begin{equation}
\Delta {\cal Z}(q,q';E) \equiv {\cal Z}(q,q';E) - {\cal Z}(0,q';E)\ .  \label{eq:3.4}
\end{equation}
This equation (albeit in different notation) was first derived
by Hammer and Mehen~\cite{HM00}.

In eq.~(\ref{eq:3.3}) we have an integral equation in which the kernel
goes to zero faster as $q''\rightarrow \infty$ than does that of the
original integral equation. As a result we hope for a unique solution
to Eq.~(\ref{eq:3.3}), even if Eq.~(\ref{eq:3.1}) does not admit a
unique solution. To establish this we need to prove that the amplitude
$K(q,0;\epsilon_d)$ is independent of the cut-off $\Lambda$,
\textit{i.e.} the solution is renormalization group invariant. Here we
proceed as above, and differentiate the subtracted equation
(\ref{eq:3.3}) with respect to the cut-off $\Lambda$, to obtain
\begin{eqnarray}
\Lambda\frac{\partial K(q,0;\epsilon_d)}{\partial \Lambda} &=&
-\frac{4m}{3}\,\Lambda \,\Delta
{\cal Z}(q,\Lambda;\epsilon_d)\,S(\Lambda;\epsilon_d)\,K(\Lambda,0;\epsilon_d)
\nonumber \\ &&-\frac{4m}{3}\int\limits_0^\Lambda dq''\,\Delta
{\cal Z}(q,q'';\epsilon_d)\, S(q'';\epsilon_d)\,\Lambda\frac{\partial
K(q'',0;\epsilon_d)}{\partial \Lambda} \ .  \label{eq:3.5}
\end{eqnarray}

Once again, we consider $\gamma \ll q \ll \Lambda$, and in this
regime:
\begin{equation}
\Delta {\cal Z}(q,\Lambda;\epsilon_d) \sim \frac{q^2}{\Lambda^4}; \qquad
S(\epsilon_d;\Lambda) \sim \Lambda;
\end{equation}
the inhomogeneous term in Eq.~(\ref{eq:3.5}) goes to zero as
$q^2/\Lambda^3$ for large $\Lambda$.  Therefore, once again, in the
limit $\Lambda\rightarrow\infty$ Eq.~(\ref{eq:3.5}) is a homogeneous
equation of the form
\begin{equation}
\Lambda\frac{\partial K(q,0;\epsilon_d)}{\partial \Lambda} = 
-\frac{4m}{3}\int\limits_0^\Lambda dq''\,\Delta {\cal Z}(q,q'';\epsilon_d)\,
S(q'';\epsilon_d)\,\Lambda\frac{\partial K(q'',0;\epsilon_d)}{\partial \Lambda}
\ .                                    \label{eq:3.6}
\end{equation}

\begin{figure}[h,b,t,p]
\centering\includegraphics[scale=0.5]{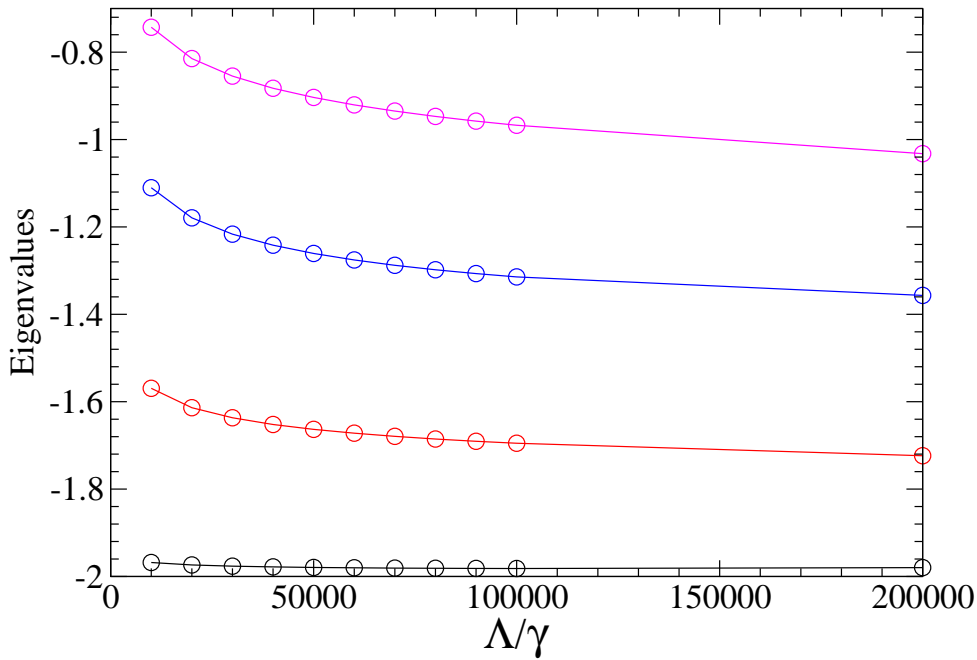}
\caption{Eigenvalues of the kernel of the subtracted integral equation 
(\ref{eq:2.17}) for various different cutoffs, for the case $a_3 \gamma=-2$.}
\label{fig-subtevals}
\end{figure}

It is now easy to show that the kernel of Eq.~(\ref{eq:3.6}) is
negative definite. Thus, no matter how large we make $\Lambda$ no
eigenvalues of one can appear. This is demonstrated numerically in
Fig.~\ref{fig-subtevals} where we plot the eigenvalues of the
subtracted integral equation's kernel as a function of
$\Lambda/\gamma$. (Here we have chosen $a_3 \gamma=-2$.)  For the
subtracted case, the kernel of the homogeneous equation,
Eq.~(\ref{eq:3.6}), has no eigenvalue close to one, thus there are no
solutions to Eq.~(\ref{eq:3.6}) and that Eq.~(\ref{eq:2.20}) is
satisfied, \textit{i.e.} the amplitude $K(q,0,\epsilon_d)$ is
independent of the cut-off $\Lambda$---up to corrections of
$O(\frac{q^2}{\Lambda^3})$ and $O(\frac{\gamma^2}{\Lambda^3})$.

\begin{figure}[h,b,t,p]
\centering\includegraphics[scale=0.75]{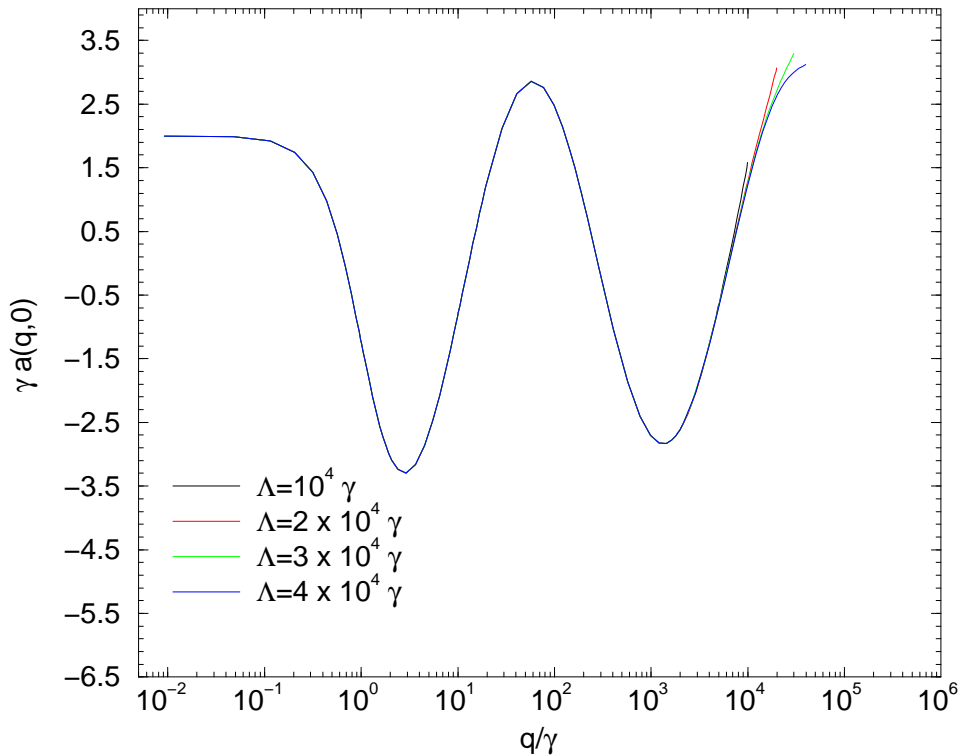}
\caption{Results for $a(q,0)$ (see Eq.~(\ref{eq:aeq}) for
definition) when various different cutoffs are imposed on the
subtracted integral equation (\ref{eq:2.17}).  Here $a_3 \gamma$ was
chosen to be -2.}
\label{fig-Kzeroensubt}
\end{figure}

In Fig.~\ref{fig-Kzeroensubt}, we present, for several values of the
cut-off $\Lambda$, the half-off-shell $K$-matrix at threshold
$K(q,0;\epsilon_d)$ that results from the subtracted equation. It is
clear from the results that the solution of the subtracted equation is
completely independent of $\Lambda$ in the regime $q \ll \Lambda$, as
anticipated from the RG argument above. In fact the cutoff can be
numerically taken to infinity without any difficulty at all.

Note that in the asymptotic regime $\gamma \ll q \ll \Lambda$ the
subtracted equation (\ref{eq:3.3}) still has solutions $K(q,0;\epsilon_d)$ of the form
(\ref{eq:asf}). These solutions ensure equality of the first piece of
the integral in (\ref{eq:3.3}) with the left-hand side of that
equation, $K(q,0;\epsilon_d)$. However, in contrast to the situation
of the previous subsection, the solution of the subtracted equation in
this asymptotic regime is not scale invariant. It must still obey
Eq.~(\ref{eq:3.2}), since those pieces of Eq.~(\ref{eq:3.3}) do not
disappear when a solution of the form (\ref{eq:asf}) is
inserted. Thus---unlike the case of Eq.~(\ref{eq:2.17})---the
asymptotic limit of (\ref{eq:3.3}) is enough to determine the
asymptotic phase $\delta$: $\delta$ is fixed such that (\ref{eq:3.2})
is obeyed.

Thus our subtracted equation at threshold yields unique
results for the half-off-shell amplitude without the need for
an explicit three-body force. It also confirms that
one piece of three-body experimental data is needed to 
properly renormalize the integral equation for the three-boson problem
in the zero-range limit.

\subsection{The subtracted equation at any energy}\label{sec3.2}

The above analysis was restricted to the amplitude at threshold and
established that the solution of the subtracted equation is
unique. The question now is: can we get the amplitude at any energy
without any further subtractions? In other words: can we use the
half-off-shell amplitude at one energy and the original 
equation (\ref{eq:2.19}) to obtain a RG-well-behaved $K_\Lambda$
at all energies?

To answer this, we need to write the on-shell amplitude at energy $E$
in terms of the solution of the half off-shell amplitude at
threshold. We do this in two stages. Rewriting Eq.~(\ref{eq:3.3}) as
\begin{equation}   
K(q,0;\epsilon_d) = K(0,0;\epsilon_d) + \Delta[{\cal Z}](q,0;\epsilon_d) -
\frac{4m}{3}\int\limits_0^\Lambda dq'' 
    \,\Delta[{\cal Z}](q,q'';\epsilon_d)\,S(q'';\epsilon_d)\,K(q'',0;\epsilon_d)\ , \label{eq:3.8}
\end{equation}
and having determined the half-off-shell amplitude at threshold, we first
need to determine the full-off-shell amplitude at threshold,
\textit{i.e.} $K(q,q';\epsilon_d)$. Before subtraction $K(q,q';\epsilon_d)$
satisfies the equation:
\begin{equation}    
K(q,q';\epsilon_d) = {\cal Z}(q,q';\epsilon_d)
-\frac{4m}{3}\int\limits_0^\Lambda dq'' \,
{\cal Z}(q,q'';\epsilon_d)\,S(q'';\epsilon_d)\,K(q'',q';\epsilon_d) \ ,
\label{eq:3.9}
\end{equation}
which has the original badly-behaved kernel of
Eq.~(\ref{eq:2.19}). So, again we need to perform a subtractive
renormalization.

Since ${\cal Z}(q,q';E) = {\cal Z}(q',q;E)$, we have that
$K(0,q;\epsilon_d)=K(q,0;\epsilon_d)$. But we know that $K(0,q';E)$,
also satisfies the equation
\begin{equation}
K(0,q';\epsilon_d) = {\cal Z}(0,q';\epsilon_d) -\frac{4m}{3}
\int\limits_0^\Lambda dq'' {\cal
Z}(0,q'';\epsilon_d)\,S(q'';\epsilon_d)\,K(q'',q;\epsilon_d)\ .
\label{eq:3.10}
\end{equation}
We can now subtract this equation from the equation for the full
off-shell amplitude---Eq.~(\ref{eq:3.9})---to get:
\begin{equation}
K(q,q';\epsilon_d) = K(0,q';\epsilon_d)
+\Delta[{\cal Z}](q,q';\epsilon_d) -\frac{4m}{3} \int\limits_0^\Lambda
dq'' \,\Delta[{\cal Z}](q,q'';\epsilon_d)\,S(q'';\epsilon_d)\,
K(q'',q';\epsilon_d) \ . \label{eq:3.11}
\end{equation} 
This equation has the same kernel as Eq.~(\ref{eq:3.8}), and given
that we have already determined
$K(0,q';\epsilon_d)=K(q',0;\epsilon_d)$, we can now determine the full
off-shell amplitude at the elastic threshold. Numerical solution
indeed confirms that the solution of Eq.~(\ref{eq:3.11}) is cutoff
independent, and that the limit $\Lambda \rightarrow \infty$ can be
taken. The resulting $K(q,q';\epsilon_d)$ is also, by construction,
real and symmetric.  In this way we have established that the full,
off-shell, amplitude at threshold can be determined with one
subtraction, and therefore, given $a_3$, we know the amplitude
$K(\epsilon_d)=X(\epsilon_d)$.

To derive the renormalized equation at any energy $E$ for the amplitude
$X(E)$, we need to write the boson-dimer equation at the energy $E$,
\textit{i.e.}
\begin{equation}
X(E) = {\cal Z}(E) + {\cal Z}(E)\,\tau(E)\,X(E) \ ,           \label{eq:3.12}
\end{equation}
and the threshold equation
\begin{equation}
X(\epsilon_d) = {\cal Z}(\epsilon_d) + {\cal Z}(\epsilon_d)\,\tau(\epsilon_d)\,X(\epsilon_d)\ .  
                                                        \label{eq:3.13}
\end{equation}
(Note that we will manipulate the equations for $X$, but the same
manipulations can equally well be done with $K$.)
These two equations can be written as:
\begin{eqnarray}
    {\cal Z}^{-1}(E) &=& X^{-1}(E) + \tau(E) \label{eq:3.14}\\
    {\cal Z}^{-1}(\epsilon_d) &=& X^{-1}(\epsilon_d) + \tau(\epsilon_d) \ .
    \label{eq:3.15}
\end{eqnarray}
We now subtract
    Eq.~(\ref{eq:3.15}) from Eq.~(\ref{eq:3.14}) with the 
result
    that
\begin{eqnarray}
 \delta[\tau] &\equiv&
 \tau(E) -
    \tau(\epsilon_d) = \left[{\cal Z}^{-1}(E)-{\cal Z}^{-1}(\epsilon_d)\right]
 -
    \left[X^{-1}(E)-X^{-1}(\epsilon_d)\right]\nonumber\\
 &=&
    {\cal Z}^{-1}(\epsilon_d)\left[{\cal Z}(\epsilon_d)-{\cal Z}(E)\right]{\cal Z}^{-1}(E)
    -X^{-1}(\epsilon_d)\left[X(\epsilon_d)-X(E)\right]X^{-1}(E)
 \ .
    \label{eq:3.16}
\end{eqnarray}
Multiply this equation from the
    left by $X(\epsilon_d)$ and from the right by 
$X(E)$, we
    get
\begin{eqnarray}
 X(E) &=& X(\epsilon_d) +
    [1+X(\epsilon_d)\,\tau(\epsilon_d)]\delta[{\cal Z}] +
    X(\epsilon_d)\delta[\tau]\,X(E)
 \nonumber \\
    &&+[1+X(\epsilon_d)\,\tau(\epsilon_d)]\,\delta[{\cal Z}]\,\tau(E)\,X(E)\
    , \label{eq:14}
\end{eqnarray}
where
\begin{equation}
    \delta[{\cal Z}] = {\cal Z}(E) - {\cal Z}(\epsilon_d) \ .
    \label{eq:3.17}
\end{equation}
All integrals in the above
    equation have sufficient ultraviolet decay to be finite, with the
    possible exception of
    $X(\epsilon_d)\,\tau(\epsilon_d)\,\delta[{\cal Z}]\,\tau(E)\,X(E)$ which
    is a double integral.

We now can write the above operator equation as an integral equation
for the amplitude at a given energy $E$ in terms of the
fully-off-shell amplitude at threshold, $X(q',q;\epsilon_d)$, as input:
\begin{equation}
X(q,k;E) = X(q,k;\epsilon_d) + B(q,k;E) +
\int\limits_0^\Lambda\,dq'\,q'^{2}\ 
 Y(q,q';E)\ X(q',k;E) \ ,
\label{eq:3.18}
\end{equation}
Where the second inhomogeneous term
is
\begin{equation}
 B(q,k;E) = \delta[{\cal Z}](q,k;E) +
\int\limits_0^\Lambda\,dq''\,
 q''^{2}\ X(q,q'';\epsilon_d)\
\tau(\epsilon_d;q'')\ \delta[{\cal Z}](q'',k;E)\ 
 ,
\label{eq:3.19}
\end{equation}
with
\begin{equation}
\delta[{\cal Z}](q,k;E) = {\cal Z}(q,k;E) - {\cal Z}(q,k;\epsilon_d)\ ,
\label{eq:3.20}
\end{equation}
and $\tau(E;q)\equiv \tau(E-\frac{3q^2}{4m})$.
Meanwhile the kernel of the integral
equation is given by
\begin{eqnarray}
 Y(q,q';E) &=&
X(q,q';\epsilon_d)\ \delta[\tau](E;q')
 + \delta[{\cal Z}](q,q';E)\
\tau(E;q') \nonumber \\&&
\qquad+\int\limits_0^\Lambda\,dq''\,q''^{2}\ X(q,q'';\epsilon_d)\ 
\tau(\epsilon_d;q'')\ \delta[{\cal Z}](q'',q';E)\ \tau(E;q')
 \nonumber \\
&=& X(q,q';\epsilon_d)\ \delta[\tau](E;q') + B(q,q';E)\ \tau(E;q)\ ,
\label{eq:3.21}
\end{eqnarray}
with
\begin{equation}
\delta[\tau](E;q) = \tau(E;q)-\tau(\epsilon_d;q) \
.\label{eq:3.22}
\end{equation}

In this way we can determine the half off-shell and from it, the on-shell
amplitude, at any energy given the on-shell amplitude at one energy where the
subtractive renormalization is done. Note that if $\delta {\cal Z}=0$, i.e.
the ``potential'' for the scattering equation is energy-independent,
then $B=0$ and $Y(E)=X(\epsilon_d) \delta[\tau](E)$.

\begin{figure}[h,b,t,p]
\vspace*{1.3cm}
\centering\includegraphics[scale=0.55]{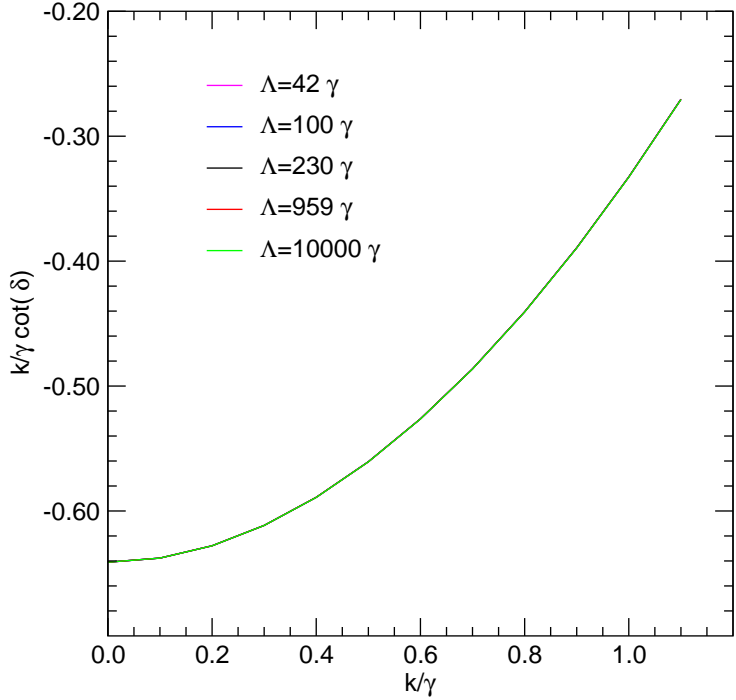}
\caption{Phase shifts for boson-dimer scattering in the case $a_3
\gamma=1.56$, for a number of different cutoffs. The curves are
completely indistinguishable.}
\label{fig-bosons}
\end{figure}

To test our procedure, we have calculated the boson-dimer phase
shifts, for the case $a_3 \gamma=1.56$. This value was chosen since
models of the helium-4 dimer suggest a ratio of three-body and
two-body scattering lengths of this size~\cite{Bd99A,Bd99B}. In
Figure~\ref{fig-bosons} we plot $k \cot (\delta)$ against $k$ (in
units of $\gamma$) for five different cutoffs. Our results agree
exactly with those reported in Ref.~\cite{Bd99A,Bd99B}. In contrast to
the figure presented in Ref.~\cite{Bd99B}, we see absolutely no cutoff
dependence whatsoever in our results. No explicit three-body force is
required to perform this renormalization.

\section{Neutron-deuteron scattering in the doublet channel}\label{sec4}

In the last section we formally developed the procedure for
calculating the amplitude for 1+2-scattering in the three-boson system
at any energy, having renormalized the equation at threshold using the
boson-dimer scattering length as input experimental data.  The final
results were identical for different cut-offs $\Lambda$.  So far in
our analysis we considered the three-boson problem in order to avoid
the additional complication of coupled-channel integral
equations. However, in order to establish the ability of the
once-subtracted equations in EFT to reproduce experimental nd
scattering data, we need to introduce the spin and isospin dependence
of the nd scattering problem. The tensor interaction in the $NN$
system is only manifest at higher order in the EFT without explicit
pions, and in this Section we will restrict our analysis to lowest
order in this EFT, known as EFT$(\not \! \pi)$, thus here we need only
include the $^1$S$_0$ and $^3$S$_1$ nucleon-nucleon channels. Since
the nucleon-nucleon interaction in the $^1$S$_0$ has anti-bound state,
we can still write a dimer-like propagator in this channel, but now
the subtraction point must be the energy of the anti-bound state. As a
result we write the quasi-deuteron propagator as
\begin{equation}
\tau_\alpha(E) =\frac{S_\alpha(E)}{E-\epsilon_\alpha}\quad\mbox{with}\quad
\alpha = s, t\ ,                         \label{eq:4.1}
\end{equation}
where $s$ and $t$ stand for the spin singlet and spin triplet nucleon-nucleon channels respectively, and:
\begin{eqnarray}
\epsilon_s=-\frac{1}{m a_s^2}; &\qquad& a_s^{-1}=-7.88~{\rm MeV},\\
\epsilon_d=-\frac{\gamma^2}{m}; &\qquad&  \gamma=45.71~{\rm MeV}.
\end{eqnarray}

The nd equations in the doublet channel are now a set of two coupled
integral equations in which the initial channel has the deuteron in
the triplet ($t$), while intermediate states can have either a singlet
($s$) or triplet ($t$) $NN$ pair with a spectator nucleon, all coupled to
spin and iso-spin one half. These equations take the form
\begin{eqnarray}
K_{tt}(q,q';E) &=& {\cal Z}_{tt}(q,q';E) + {\cal P} \int\limits_0^\Lambda\
dq''\,q''^2
\ {\cal Z}_{tt}(q,q'';E)\,\tau_t\left(E-\frac{q''^2}{2\nu}\right)\
K_{tt}(q'',q';E)
\nonumber \\
&&\qquad + {\cal P} \int\limits_0^\Lambda\
dq''\,q''^2\ {\cal Z}_{ts}(q,q'';E)\,
\tau_s\left(E-\frac{q''^2}{2\nu}\right)\
K_{st}(q'',q';E)\label{eq:4.2}\\
K_{st}(q,q';E) &=& {\cal Z}_{st}(q,q';E) +
{\cal P} \int\limits_0^\Lambda\ dq''\,q''^2
\
{\cal Z}_{st}(q,q'';E)\,\tau_t\left(E-\frac{q''^2}{2\nu}\right)\
K_{tt}(q'',q';E)
\nonumber \\
&&\qquad + {\cal P} \int\limits_0^\Lambda\
dq''\,q''^2\ {\cal Z}_{ss}(q,q'';E)\,
\tau_s\left(E-\frac{q''^2}{2\nu}\right)\
K_{st}(q'',q';E)\ .\label{eq:4.3}
\end{eqnarray} 
where $\nu =
\frac{2}{3}m$ is the reduced mass for the nd system, and the Born
amplitude ${\cal Z}_{\alpha\beta}$ is given by
\begin{equation}
{\cal Z}_{\alpha\beta}(q,q';E) = -\lambda_{\alpha\beta}\ \frac{m}{qq'}\
\ln\left\{\frac{q^2+q'^2+qq'-mE}{q^2+q'^2-qq'-mE}\right\}\ ,
\label{eq:4.4}
\end{equation} 
with the spin iso-spin factor matrix ${\bf \lambda}$ given by:
\begin{equation}
{\bf \lambda} = \frac{1}{4}
\left(\begin{array}{cc}
1 & -3 \\
-3  & 1\end{array}\right)    \ .   \label{eq:4.5}
\end{equation}

It is not immediately apparent that the kernel of the coupled integral
equations (\ref{eq:4.2}) and (\ref{eq:4.3}) has the same problems as
that of (\ref{eq:2.19}). By taking linear combinations of
(\ref{eq:4.2}) and (\ref{eq:4.3}) and looking in the asymptotic region
we can perform an analysis akin to that used for
Eq.~(\ref{eq:2.19})~\cite{Bd00}. This shows that one subtraction is
required to render the system (\ref{eq:4.2})--(\ref{eq:4.3})
well-behaved. Otherwise this kernel too, has eigenvalues which cross
one as $\Lambda$ is increased, and the RG-evolution of $K_{tt}$ at low
momenta will not be smooth.

In this case the $K_{tt}(0,0;\epsilon_d)$, as given by Eq.~(\ref{eq:thold}),
with $a_3$ the doublet scattering length, is chosen for the subtraction. 
Experimentally~\cite{Di71}:
\begin{equation}
a_3=0.65 \pm 0.04~{\rm fm}.
\label{eq:4.6}
\end{equation}
We adopt the central value for $a_3$.

After the subtraction is performed the equations for the half-off-shell
threshold amplitude become:
\begin{eqnarray}
K_{tt}(q,0;\epsilon_d) &=& K_{tt}(0,0;\epsilon_d) + 
\Delta {\cal Z}_{tt}(q,0;\epsilon_d)\nonumber\\
&&\qquad +  \int\limits_0^\Lambda\
dq''\,q''^2 \ \Delta {\cal Z}_{tt}(q,q'';E)\,\tau_t\left(\epsilon_d-\frac{q''^2}{2\nu}\right)\
K_{tt}(q'',0;\epsilon_d)\nonumber \\
&&\qquad +  \int\limits_0^\Lambda\
dq''\,q''^2\ \Delta {\cal Z}_{ts}(q,q'';\epsilon_d)\,
\tau_s\left(\epsilon_d-\frac{q''^2}{2\nu}\right)\
K_{st}(q'',0;\epsilon_d)\label{eq:4.7}\\
K_{st}(q,0;\epsilon_d) &=& K_{tt}(0,0;\epsilon_d) + [{\cal Z}_{st}(q,0;\epsilon_d) 
-{\cal Z}_{tt}(0,0;\epsilon_d)]\nonumber\\
&& \qquad +
 \int\limits_0^\Lambda\ dq''\,q''^2
\
[{\cal Z}_{st}(q,q'';\epsilon_d) - {\cal Z}_{tt}(0,q'';\epsilon_d)]
\,\tau_t\left(\epsilon_d-\frac{q''^2}{2\nu}\right)\
K_{tt}(q'',0;\epsilon_d)
\nonumber \\
&&\qquad +  \int\limits_0^\Lambda\
dq''\,q''^2\ [{\cal Z}_{ss}(q,q'';\epsilon_d) - {\cal Z}_{ts}(0,q'';\epsilon_d)]
\,
\tau_s\left(\epsilon_d-\frac{q''^2}{2\nu}\right)\
K_{st}(q'',0;\epsilon_d)\ .\nonumber\\
&&\qquad\label{eq:4.8}
\end{eqnarray} 

Once these equations have been solved for $K_{tt}(q,0;\epsilon_d)$ and
$K_{st}(q,0;\epsilon_d)$ we can demand:
\begin{equation}
K_{tt}(q,0;\epsilon_d)=K_{tt}(0,q;\epsilon_d); \qquad
K_{ts}(0,q;\epsilon_d)=K_{st}(q,0;\epsilon_d),
\label{eq:4.9}
\end{equation}
and so arrive at two sets of two coupled equations apiece. These four
equations determine the fully-off-shell threshold nd scattering amplitude.
The first pair is:
\begin{eqnarray}
K_{tt}(q,q';\epsilon_d) &=& K_{tt}(0,q';\epsilon_d) + 
\Delta {\cal Z}_{tt}(q,q';\epsilon_d)\nonumber\\
&&\qquad +  \int\limits_0^\Lambda\
dq''\,q''^2 \ \Delta {\cal Z}_{tt}(q,q'';\epsilon_d)\,
\tau_t\left(\epsilon_d-\frac{q''^2}{2\nu}\right)\
K_{tt}(q'',q';\epsilon_d)
\nonumber \\
&&\qquad +  \int\limits_0^\Lambda\
dq''\,q''^2\ \Delta {\cal Z}_{ts}(q,q'';\epsilon_d)\,
\tau_s\left(\epsilon_d-\frac{q''^2}{2\nu}\right)\
K_{st}(q'',q';\epsilon_d)\label{eq:4.10}\\
K_{st}(q,q';\epsilon_d) &=& K_{tt}(0,q';\epsilon_d) + [{\cal Z}_{st}(q,q';\epsilon_d) 
-{\cal Z}_{tt}(0,q';\epsilon_d)]\nonumber\\
&& \qquad +
 \int\limits_0^\Lambda\ dq''\,q''^2
\
[{\cal Z}_{st}(q,q'';\epsilon_d) - {\cal Z}_{tt}(0,q'';\epsilon_d)]
\,\tau_t\left(\epsilon_d-\frac{q''^2}{2\nu}\right)\
K_{tt}(q'',q';\epsilon_d)
\nonumber \\
&&\qquad +  \int\limits_0^\Lambda\
dq''\,q''^2\ [{\cal Z}_{ss}(q,q'';\epsilon_d) - {\cal Z}_{ts}(0,q'';\epsilon_d)]
\,
\tau_s\left(\epsilon_d-\frac{q''^2}{2\nu}\right)\
K_{st}(q'',q';\epsilon_d)\ ,\nonumber\\
\label{eq:4.11}
\end{eqnarray} 
which have exactly the same kernel as (\ref{eq:4.7}) and (\ref{eq:4.8}),
but different driving terms.

The second set of subtracted equations describes the (unphysical)
amplitudes $K_{ts}$ and $K_{ss}$ at threshold. The unsubtracted
versions of these equations are given by a simple extension of
Eqs.~(\ref{eq:4.2}) and (\ref{eq:4.3}). After subtraction the
equations are:
\begin{eqnarray}
K_{ts}(q,q';\epsilon_d)&=&K_{ts}(0,q';\epsilon_d) + \Delta {\cal
Z}_{ts}(q,q';\epsilon_d)\nonumber\\ &&\qquad + \int\limits_0^\Lambda
dq'' \, q''^2 \Delta {\cal Z}_{tt} (q,q'';\epsilon_d)
\tau_t\left(\epsilon_d - \frac{q''^2}{2 \nu}\right)
K_{ts}(q'',q';\epsilon_d) \nonumber\\ && \qquad +
\int\limits_0^\Lambda dq'' \, q''^2 \Delta {\cal Z}_{ts}
(q,q'';\epsilon_d) \tau_s\left(\epsilon_d - \frac{q''^2}{2 \nu}\right) 
K_{ss}(q'',q';\epsilon_d) \label{eq:4.12}\\ 
K_{ss}(q,q';\epsilon_d)&=&K_{ts}(0,q';\epsilon_d) +
{\cal Z}_{ss}(q,q';\epsilon_d) - {\cal
Z}_{ts}(0,q';\epsilon_d)\nonumber\\ &&\qquad + \int\limits_0^\Lambda
dq'' \, q''^2 [{\cal Z}_{st} (q,q'';\epsilon_d) - {\cal
Z}_{tt}(0,q'';\epsilon_d)] \tau_t\left(\epsilon_d - \frac{q''^2}{2
\nu}\right) K_{ts}(q'',q';\epsilon_d) \nonumber\\ && \qquad +
\int\limits_0^\Lambda dq'' \, q''^2 [{\cal Z}_{ss} (q,q'';\epsilon_d)
- {\cal Z}_{ts}(q,q'';\epsilon_d)] \tau_s\left(\epsilon_d -
\frac{q''^2}{2 \nu}\right) K_{ss}(q'',q';\epsilon_d). 
\nonumber\\
\label{eq:4.13}
\end{eqnarray}
Note that imposing (\ref{eq:4.9}) to perform the subtraction on the
set of four original integral equations (written in matrix form in 
Appendix~\ref{app-connect}) leads to a symmetric result for
the 2 $\times$ 2 matrix form of the threshold amplitude.

Now we write the original, unsubtracted, equations in operator form, as:
\begin{equation}
{\bf K}(E)={\bf {\cal Z}}(E) + {\bf {\cal Z}}(E) \, {\bf \tau}(E) \, {\bf K}(E),
\end{equation}
with ${\bf K}$ the 2 $\times$ 2 matrix:
\begin{equation}
{\bf K} \equiv \left(\begin{array}{cc}
               K_{tt} & K_{ts} \\
	       K_{st} & K_{ss} 
		\end{array} \right);
\end{equation}
\begin{equation}
{\bf \tau}(E;q'') \equiv \left(\begin{array}{cc}
               \tau_t\left(E - \frac{q''^2}{2 \nu}\right) & 0 \\
	        0     & \tau_s \left(E - \frac{q''^2}{2\nu}\right) 
		\end{array} \right).
\end{equation}
and ${\bf {\cal Z}}$ the 2 $\times$ 2 matrix defined by Eq.~(\ref{eq:4.4}).

We can then perform the formal manipulations that lead to
Eqs.~(\ref{eq:3.18})--(\ref{eq:3.22}), except that now
all quantities are 2 $\times$ 2 matrices in channel space, and thus the
final integral equation to be solved is, in matrix form, but 
with the momentum-dependence made explicit:
\begin{equation}
{\bf X}(q,k;E)={\bf X}(q,k;\epsilon_d) + {\bf B}(q,k;E) + \int
\limits_0^\Lambda dq'' \, q''^2 \, {\bf Y} (q,q'';E) \, {\bf X}(q'',k;E),
\end{equation}
with:
\begin{eqnarray}
{\bf B}(E)&=&\delta [{\bf {\cal Z}}] + {\bf X}(\epsilon_d) {\bf \tau}(\epsilon_d) 
\delta [{\bf {\cal Z}}] \\
{\bf Y}(E)&=&{\bf X}(\epsilon_d) \delta[\tau] + {\bf B}(E) {\bf \tau}(E),
\end{eqnarray}
where the meaning of the energy-difference operator $\delta$ is exactly
as in the boson case of the previous section.

\begin{figure}[h,b,t,p]
\centering\includegraphics[scale=0.55]{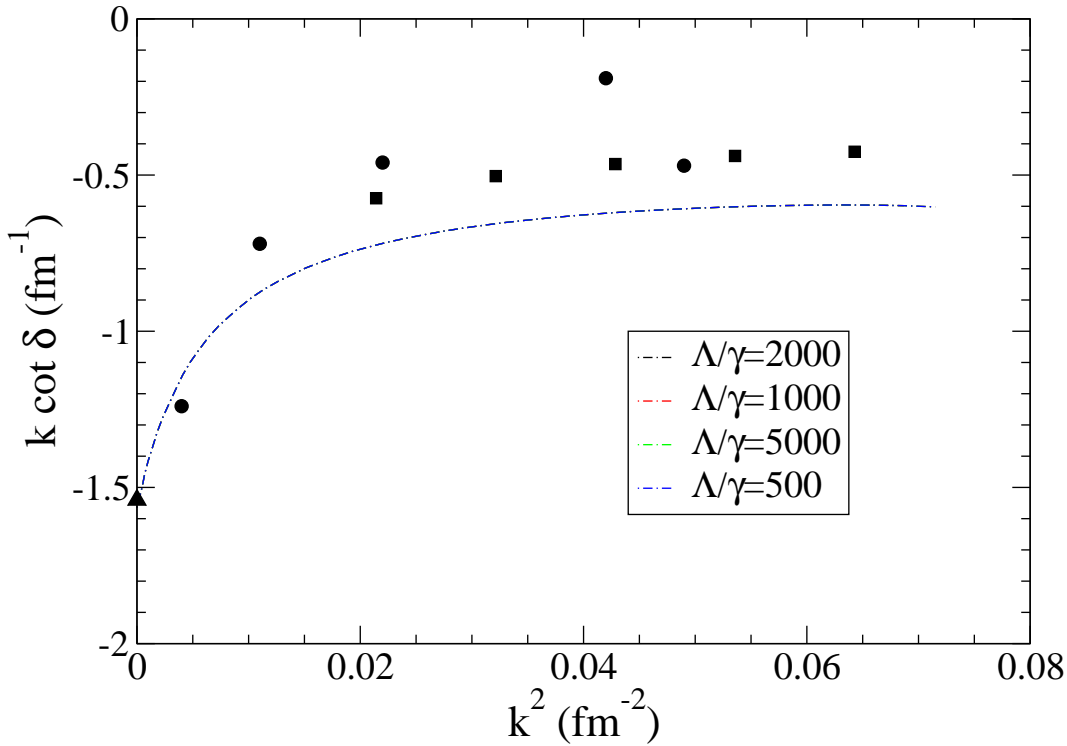}
\caption{Phase shifts for neutron-deuteron scattering at low energies,
at leading order in the nucleon-nucleon EFT without explicit pions,
for a variety of different cutoffs. The curves for different cutoffs
are indistinguishable. The triangle is the central value of the
scattering length measurement of Ref.~\cite{Di71}. The circles are the
results of the van Oers-Seagrave phase-shift analysis~\cite{vOS67},
and the squares represent a recent calculation of these phase shifts
by Kievsky et al.~\cite{Ki96}.}
\label{fig-LOdoublet}
\end{figure}

Applying these equations to scattering in the nd doublet channel below
the nnp breakup threshold yields the phase shifts shown in
Fig.~\ref{fig-LOdoublet}. At almost all energies shown these agree
with the leading-order results published in Ref.~\cite{HM01} at the
1\% level. Once again, there is no cutoff dependence, once the doublet
scattering length is used to subtractively renormalize the
equations. Also shown are the results of a phase-shift
analysis~\cite{vOS67}, and the results of a calculation using the AV18
$NN$ and UIX $NNN$ potential~\cite{Ki96}.

\section{The nd doublet channel beyond leading order}\label{sec5}

In this section we discuss calculations of nd doublet scattering which
go beyond the leading-order calculation of the previous section.  That
computation employed the Lagrangian (\ref{eq:2.1}), extended
to the $NN$ system. This Lagrangian
is equivalent to~\cite{BG00}:
\begin{equation}
{\cal L}^{(0)}={\cal L} = N^\dag \left(i \partial_0 +
\frac{\nabla^2}{2M}\right)N
- \sum_{\alpha=s,t} {}^{\not\pi} C_{0,-1}^{(\alpha)}
(N^T P^\alpha_a N)^\dagger (N^T P^\alpha_a N) 
\label{eq:5.1}
\end{equation}
with $P^\alpha_a$ the spin-isopsin projector which restricts the interactions
to the $^3$S$_1$ or $^1$S$_0$ channel, as appropriate:
\begin{equation}
P^s_a=\frac{1}{\sqrt{8}} \sigma_2 \, \tau_2 \tau_a; \qquad
P^t_a=\frac{1}{\sqrt{8}} \sigma_2 \sigma_a \, \tau_2,
\label{eq:5.2}
\end{equation}
and ${}^{\not \pi} C_{0,-1}^{(\alpha)}$ is the leading-order contact
interaction in these channels. The subscript $_{0,-1}$ on this
coefficient indicates that it appears in front of an interaction which
has no derivatives, but that it scales as $Q^{-1}$, with the
enhancement over its naive dimensional analsysis scaling being due to
the presence of the unnaturally-large scattering lengths in the two-body
system~\cite{vK97,vK98,Ka98A,Ka98B,Bi99}.

The higher-order calculation we report on in this section requires the
insertion of higher-derivative four-nucleon operators. The analysis of
Refs.~\cite{vK98,Ka98A,Ka98B,Bi99} indicates that the first
additional piece of the EFT Lagrangian which must be considered
is~\cite{Be00}:
\begin{equation}
{\cal L}^{(1)}=- \sum_{\alpha=s,t} {}^{\not\pi} C_{0,0}^{(\alpha)}
(N^T P^\alpha_a N)^\dagger (N^T P^\alpha_a N) - {}^{\not\pi}
C_2^{(\alpha)} \frac{1}{2}\left[(N^T P_a^\alpha N)^\dagger (N^T {\cal
O}^{2,\alpha}_a N) + {\rm h.c.}\right]
\label{eq:5.3}
\end{equation}
and the Hermitian, two-derivative three-component, operator ${\cal
O}^{2,\alpha}$ is defined by:
\begin{equation}
{\cal O}^{2,\alpha}_a=-\frac{1}{4} \left[P^\alpha_a \overrightarrow
{\bf \nabla}^2 +\overleftarrow {\bf \nabla}^2 P^\alpha_a - 2
\overleftarrow {\bf \nabla} P^\alpha_a \overrightarrow {\bf
\nabla}\right].
\label{eq:5.4}
\end{equation}
Here the effect of the two-derivative operators on the $NN$ amplitude
is suppressed by one power of the small parameter $\gamma R$ ($R/a_s$ in
the $^1$S$_0$ case) relative to the leading-order EFT amplitude. Also
appearing in ${\cal L}^{(1)}$ is a small correction to $C_0$, denoted
by ${}^{\not \pi} C_{0,0}$: ``small'' because ${}^{\not \pi}
C_{0,0}$ is down by $\gamma R$ relative to ${}^{\not \pi}
C_{0,-1}$~\cite{vK98,Ka98A,Ka98B,Bi99}.

Thus the effects of the terms in ${\cal L}^{(1)}$ on the $NN$
amplitude can be calculated in perturbation theory. ${}^{\not \pi}
C_2^{({}^3S_1)}$ can then be chosen so as to reproduce the asymptotic S-state
normalization of deuterium~\cite{Ph99}, and ${}^{\not \pi} C_{0,0}^{({}^3S_1)}$
adjusted in such a way that double-pole term which would otherwise
appear in the $NN$ EFT amplitude is removed. This produces a
next-to-leading order (NLO) $^3$S$_1$ $NN$ amplitude~\cite{Be00}:
\begin{equation}
T(p)=\frac{2}{\pi m} \left[\frac{Z_t}{\gamma + i p} - \frac{Z_t -
1}{2 \gamma} \right],
\label{eq:5.5a}
\end{equation}
with $p=\sqrt{ME}$, and where $Z_t$ is the residue of the $^3$S$_1$
T-matrix at the deuteron pole $p=i \gamma$. This amplitude is easily
seen to be a re-expanded version of the effective-range-theory
$^3$S$_1$ amplitude:
\begin{equation}
T(p)=\frac{2}{\pi m} \frac{1}{\gamma - \frac{1}{2} \rho_t(p^2 + \gamma^2)
+ i p},
\label{eq:5.5b}
\end{equation}
where $\rho_t$ is the $NN$ $^3$S$_1$ effective range, which is of
order the range of the $NN$ interaction: $\rho_t \sim R$. The
re-expansion is thus in the small parameters $\rho_t \gamma$ and
$\rho_t p$, but with $\gamma$ treated as being of the same size as
$p$.  By making such an identification we determine that:
\begin{equation}
Z_t=\frac{1}{1 - \gamma \rho_t}.
\label{eq:5.6}
\end{equation}
$Z_t$ is also related to the asymptotic S-state
normalization of deuterium, $A_S$~\cite{PC99,Ph99}:
\begin{equation}
A_S^2=2 \gamma Z_t.
\label{eq:5.7}
\end{equation}
Using the Nijmegen PSA value for $A_S$, $A_S=0.8845~{\rm
fm}^{-1/2}$~\cite{St93,deS95}, we obtain:
\begin{equation}
Z_t=1.686,
\end{equation}
which agrees with the result obtained from Eq.~(\ref{eq:5.6}) to three
significant figures.

To summarize, the coefficients in the NLO EFT Lagrangian may be chosen
such that the amplitude in the $^3$S$_1$ channel has a deuteron pole
with the experimental binding energy and the ``experimental''
asymptotic S-state normalization. Also present in the NLO $NN$ $^3$S$_1$
amplitude is a constant piece, which is proportional to $\rho_t$. Here
we wish only to assess the impact of higher-order terms on the nd
phase shifts, and thus, we will perform a partial NLO calculation of
the nd phase shifts below breakup threshold, dropping the non-pole
term in Eq.~(\ref{eq:5.4}). Work on complete higher-order calculations
within our subtractive framework is in progress, and these numerical
studies, as well as prior results by other authors~\cite{HM01,Bd02}
indicate that including the constant term of Eq.~(\ref{eq:5.4}) has
little effect on nd phase shifts below nd breakup
threshold~\footnote{In Ref.~\cite{HM01} it was argued that the
constant term actually gives zero contribution to nd phase shifts, and
so it was dropped there too. Although the contribution is not, in
fact, strictly zero, it is small, as witnessed by the good agreement
between the NLO results of Ref.~\cite{HM01} and Ref.~\cite{Bd02},
where the non-pole $^3$S$_1$ term was included in the analysis.}.

Similar results follow for the NLO $^1$S$_0$ amplitude, and there:
\begin{equation}
Z_s=\frac{1}{1-r_s/a_s},
\end{equation}
$r_s=2.73$ fm~\cite{St93} being the effective range in this channel.
This results in much smaller NLO corrections from this channel, since
$r_s/a_s$ is only of order 10\%.

Thus, to perform our (partial) next-to-leading-order calculation for
nd scattering the only changes to the amplitude which are necessary
are the multiplication of $\tau_t$ and $\tau_s$ by factors $Z_t$ and
$Z_s$. The subtractive procedure developed above is not affected by
the inclusion of these factors: the only changes necessary in the
above equations are the replacements
\begin{equation}
S_t \rightarrow S_t Z_t; \qquad S_s \rightarrow S_s Z_s.
\end{equation}

\begin{figure}[h,b,t,p]
\centering\includegraphics[scale=0.5]{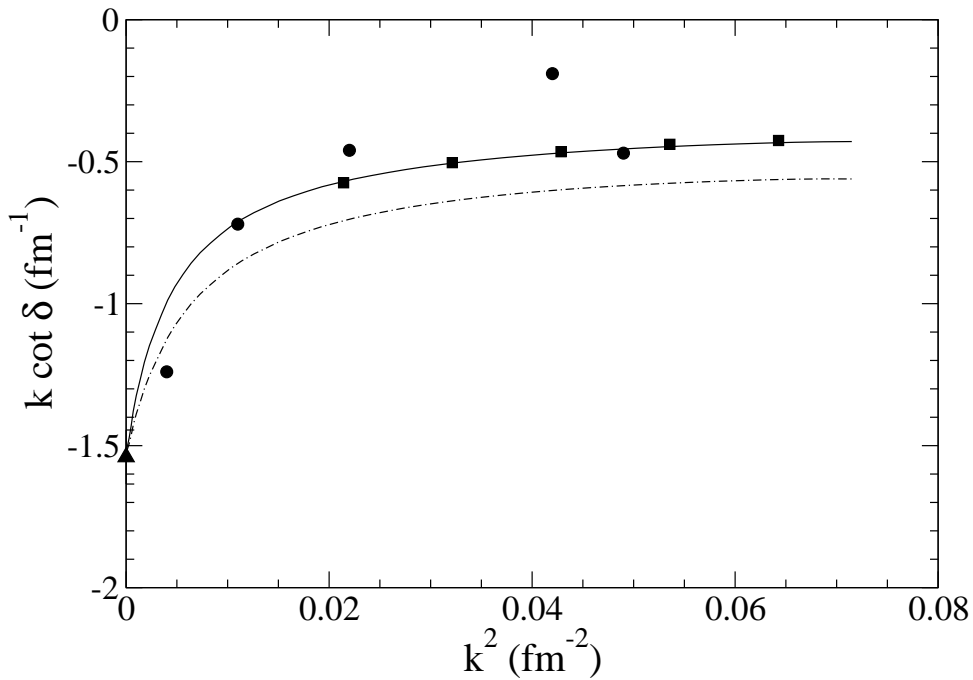}
\caption{Phase shifts for neutron-deuteron scattering at low energies,
at leading order (dash-dotted) and next-to-leading order (solid) in
the nucleon-nucleon EFT without explicit pions. The curves for
different cutoffs are indistinguishable. Data are as described
in Fig.~\ref{fig-LOdoublet}.}
\label{fig-higherorder}
\end{figure}

Making these replacements we obtain the results shown in
Fig.~\ref{fig-higherorder}. Once again the result is cutoff
independent. It agrees remarkably well with the sophisticated
potential-model calculation of Kievsky {\it et al.}~\cite{Ki96}. The
agreement with the single-energy phase-shift analysis of van Oers and
Seagrave~\cite{vOS67} is not as pleasing, but it is clear that modern
potential-model calculations do not agree with these older doublet phase
shifts either.

These results are a (partial) NLO calculation of the nd phase shifts
below breakup. They differ from those of Refs.~\cite{HM01,Bd02} since
in those works the authors chose to adjust the NLO coefficients in the
EFT Lagrangian to reproduce $\rho_t$ exactly, and so only obtained
$Z_t$ (or equivalently $A_S$) approximately. The difference between
this ``$\rho$-parameterization'' and our ``$Z$-parameterization'' is a
higher-order effect, and the magnitude of the discrepancy between the
results of Fig.~\ref{fig-higherorder} and those of
Refs.~\cite{HM01,Bd02} is consistent with an effect of order $(\rho_t
\gamma)^2$, i.e. two orders beyond leading. Comparison of our
numerical results with those of Ref.~\cite{Bd02} indicates that if we
adopt the $\rho$-parameterization the agreement is better than
1\%~\cite{Gr03}.

\section{Conclusions}\label{sec6}

The integral equation which describes 1+2-scattering in the effective
field theory with short-range interactions alone does not yield an
RG-invariant low-energy amplitude.  By performing an RG-analysis of
the integral equation for this process we traced this poor RG
behaviour to the presence of eigenvalues of order one in the kernel of
the integral equation. One subtraction removes these eigenvalues from
the spectrum of the kernel and renders it negative definite (at
threshold). Imposing Hermiticity and employing a series of resolvent
identities we can use this single subtraction at the 1+2-threshold to
generate predictions for phase shifts at finite energies.  Note that
although here we have only computed phase shifts below 1+2-breakup
threshold our subtractive technique is easily extended to include
energies above the three-body threshold. The only complication is the
technical one of dealing with the logarithmic branch cuts that appear
in the kernel of the integral equation at this energy.

The equations we have developed are equivalent to the equations of
Bedaque {\it et al.}~\cite{Bd99A,Bd99B}, and may be obtained from
those equations by algebraic manipulations. The distinguishing feature
of our formulation is that the equations are subtractively
renormalized, i.e. only physical quantities appear in them, and any
regulator can be employed. This would appear to make this formulation
especially useful for higher-order computations in the nd system. It
also provides particular emphasis to the point that -- as in the case with all bare parameters in field-theoretic Lagrangian -- the three-body
force which appears in the equations of Bedaque {\it et al.} is not an
observable.

Thus one piece of three-body experimental data is needed in order to
renormalize the three-body equations for zero-range forces. For this
piece of data we choose the 1+2 scattering length.  Its value can be
incorporated into the EFT description of the three-body system either
via a counterterm, as in Refs.~\cite{Bd99A,Bd99B,Bd00} or, as done
here, by a subtraction of the badly-behaved integral equation. The
renormalization of the equation after the inclusion of this single
piece of three-body data provides a simple, model-independent,
explanation for well-known features of the three-nucleon system such
as the Phillips line~\cite{Ph68}.  It also facilitates the
systematization of predictions made by Efimov for such
systems~\cite{Ef81,Ef85,Ef91,Ef93}.

Finally, we performed a partial treatment of next-to-leading order
corrections to the doublet nd phase shifts in the EFT. We found that
adopting coefficients in the NLO EFT($\not \!  \pi$) Lagrangian that
give the correct deuteron binding energy and asymptotic S-state
normalization results in excellent reproduction of potential-model
$S=\frac{1}{2}$ nd phase shifts below deuteron breakup threshold. Our
results suggest that---to a very good level of approximation--- these
nd phase shifts are determined by four numbers from the two-body
system, $\epsilon_d$, $A_S$, and the $^1$S$_0$ scattering length and
effective range, together with the crucial one piece of data from the
three-body system: the nd doublet scattering length.

\acknowledgments{D.~R.~P. thanks Silas Beane, Paulo Bedaque, Harald
Grie\ss hammer, and Hans-Werner Hammer for useful discussions. We also
thank Hans-Werner Hammer and Harald Grie\ss hammer for providing us
with their results, and with phase-shift data for nd scattering in the
doublet channel. D.~R.~P. is grateful for the hospitality of Flinders
University, where much of this work was done, and that of the
Institute for Nuclear Theory, where it was completed. The work of
D.~R.~P. is supported by the United States Department of Energy under
DE-FG02-93ER40756 and DE-FG02-02ER41218. The work of I.~R.~A. is
supported by the Australian Research Council.}

\appendix

\section{Spin-isospin factors for the Amado equation}
\label{app-recoupling}

In this appendix we derive the spin-isospin factors for the Amado model
for: (i)~three bosons; (ii)~nd quartet; (iii)~ nd doublet. In the
latter two cases we restrict our analysis to S-waves only.

The Amado equation can be written in operator form as
\[
X = 2Z + 2Z\,\tau\ X
\]
where
\[
Z_{\alpha\beta} = (1-\delta_{\alpha\beta})\bra g_\alpha|G_0(E)|g_\beta\ket
\]
with $G_0(E)=(E-H_0)^{-1}$. \textbf{Note:} This differs from Lovelace~\cite{Lo64} by $(-1)$ due to a different definition of $G_0$.

This $Z_{\alpha\beta}$ can be written after partial wave expansion as
\[
Z_{\alpha\beta} = \lambda_{\alpha\beta}\ \frac{1}{2}\,\int\limits_{-1}^{+1} dx
\frac{g_\alpha(p_\alpha)\ g_\beta(p_\beta)}{E - \frac{1}{m}(q_\alpha^2 +
q_\beta^2 + q_\alpha q_\beta x)}\ P_\ell(x)
\]
where $\lambda_{\alpha\beta}$ is the product of a spin factor $\Lambda^S_{\alpha\beta}$ and an isospin factor $\Lambda^I_{\alpha\beta}$, \textit{i.e.}
\[
\lambda_{\alpha\beta} = \Lambda^S_{\alpha\beta}\ \Lambda^I_{\alpha\beta}\ .
\]
with
\[
\Lambda^S_{\alpha\beta} = (-1)^{s_\beta + s_\gamma -S_\alpha+2S} \ \left[(2S_\alpha+1)(2S_\beta+1)\right]^{1/2}\ 
\left\{\begin{array}{ccc}
s_\alpha & s_\gamma & S_\beta \\ 
s_\beta & S & S_\alpha
\end{array}\right\}\ ,
\]
where $s_\alpha$, $s_\beta$ and $s_\gamma$ are the spin of the three particles, and $S_\alpha$ is the total spin of the pair $(\beta\gamma)$. This expression can also be used to calculate the isospin factor $\Lambda^I_{\alpha\beta}$.

For three boson the spins and the isospin of all three particles is zero. In this case we have only one channel, \textit{i.e.} $\lambda_{\alpha\beta}=\lambda$, and therefore the spin-isospin factor is one, \textit{i.e.}
\[
\lambda_{\alpha\beta} = 1, \qquad\mbox{three bosons}\ .
\]

For nd scattering all spins and isospins are $\frac{1}{2}$. For the quartet state we have only one channel with $S=\frac{3}{2}$ and $I=\frac{1}{2}$. The spin and isospin of the pair is $1$ and $0$ (the quantum numbers of the deuteron) respectively. In this case
\[
\Lambda^S_{\alpha\beta}=1\ ,\quad \Lambda^I_{\alpha\beta}=-\frac{1}{2}\ ,
\]
and therefore
\[
\lambda_{\alpha\beta} = -\frac{1}{2}\ ,\qquad \mbox{nd quartet}\ .
\]

Finally for the case of nd doublet we have two channels. They correspond to the pair of nucleons being in either $t=(S_\alpha=1,t_\alpha=0)$ the deuteron, or in $s=(S_\alpha=0,t_\alpha=1)$ the singlet. In this case $S=I=\frac{1}{2}$. The spin isospin factors are
\[
\Lambda^S_{tt}=-\frac{1}{2}\ ,\quad \Lambda^I_{tt}=-\frac{1}{2}\quad\Longrightarrow
\quad\lambda_{tt}=\frac{1}{4}
\]
\[
\Lambda^S_{ss}=-\frac{1}{2}\ ,\quad \Lambda^I_{ss}=-\frac{1}{2}\quad\Longrightarrow\quad
\lambda_{ss}=\frac{1}{4}
\]
and
\[
\Lambda^S_{ts}=-\frac{\sqrt{3}}{2}\ ,\quad \Lambda^I_{ts}=\frac{\sqrt{3}}{2}\quad\Longrightarrow\quad
\lambda_{ts}=-\frac{3}{4}\ ,
\]
or
\[
\lambda_{\alpha\beta}=\frac{1}{4}\left(\begin{array}{cc}
1 & -3 \\ -3 & 1
\end{array}\right),\qquad \ \mbox{nd doublet}\ .
\]

\section{Connection to equations of Bedaque et al.}

\label{app-connect}

The equation of Refs.~\cite{BvK97,Bd98,Bd99B}, is, in the case of no three-body force:
\begin{equation}
a(k,p)=\lambda M(k,p;k) + \frac{2 \lambda}{\pi} \int \limits_0^\Lambda
dq \, M(q,p;k) \frac{q^2}{q^2 - k^2 - i \eta} a (k,q),
\label{eq:A1.1}
\end{equation}
with
\begin{equation}
M(q,p;k)=\frac{4}{3} \left(\gamma + \sqrt{\frac{3 p^2}{4} - mE}\right)
\frac{1}{p q} \ln\left(\frac{q^2 + qp + p^2 - mE}{q^2 - qp + p^2 
- mE}\right),
\end{equation}
and:
\begin{equation}
\lambda=\left\{\begin{array}{ll}
		1  &\mbox{for three bosons,}\\
		-\frac{1}{2} & \mbox{for the nd quartet channel.}
		\end{array} \right.
\end{equation}
Here the relation to the phase shifts is given simply by:
\begin{equation}
\mbox{Re}\left(\frac{1}{a(k,k)}\right)=k \cot \delta.
\label{eq:A1.4}
\end{equation}

To make the connection to Eq.~(\ref{eq:2.9}) first observe that:
\begin{equation}
S(E;q)=\frac{2}{\pi} \frac{1}{m^2} \left(\gamma + \sqrt{\frac{3}{4} q^2 
- mE}\right),
\end{equation}
and then define $X(p,k:E)$ such that:
\begin{equation}
\frac{4}{3} \left(\gamma + \sqrt{\frac{3
p^2}{4} - mE}\right) X(p,k;E) \equiv -m a(k,p),
\label{eq:A1.6}
\end{equation}
we then find:
\begin{equation}
X(p,k;E)={\cal Z}(p,k;E) + \int \limits_0^\Lambda dq \, q^2 \, {\cal Z}(p,q;E) \frac{S(E;q)}{E - \frac{3 q^2}{4m} 
- \epsilon_d} \, X(q,k;E),
\label{eq:2.9redux}
\end{equation}
with ${\cal Z}(p,q;E)$ given exactly by Eq.~(\ref{eq:2.10}) above. Note, in
particular, that the homogeneous equation corresponding to
(\ref{eq:2.10}) requires no manipulation to be equivalent to that
corresponding to Eq.~(\ref{eq:A1.1}). The
relationship of $X$ to the phase shifts can be deduced from
Eq.~(\ref{eq:A1.4}) and (\ref{eq:A1.6}). It is:
\begin{equation}
\mbox{Re}\left(\frac{1}{X(k,k;E)}\right)=-\frac{8 \gamma}{3 m} \ k \cot \delta \ ,
\label{eq:A8}
\end{equation}
in agreement with Eq.~(\ref{eq:2.17}).

In the case of nd scattering in the doublet channel we begin with the
coupled equations of Ref.~\cite{Bd02}, which, again in the absence of
a three-body force term, may be written in matrix form as:
\begin{equation}
{\bf t}(p,k)={\bf V}(p,k) + \frac{2}{\pi} \int \limits_0^\Lambda dq \, q^2 \,
{\bf V}(p,q;k) \, {\bf D}(q;k) \, {\bf t}(q;k),
\end{equation}
with:
\begin{equation}
{\bf t} \equiv \left(\begin{array}{cc}
               t_{tt} & t_{ts} \\
	       t_{st} & t_{ss} 
		\end{array} \right);\qquad
{\bf D} \equiv \left(\begin{array}{cc}
               D_t & 0 \\
	       0 & D_s
		\end{array} \right),
\end{equation}
and
\begin{equation}
{\bf V}\equiv
\frac{1}{4pq} \ln\left[\frac{p^2 +q^2+pq-mE}{p^2+q^2-pq-mE}\right] \times
\left(\begin{array}{cc}
                     1   & 3  \\
                     3  &  1 
		        \end{array} \right).
\end{equation}
Here, to leading order in $\gamma R$,
\begin{eqnarray}
D_t(q;k)&=&\frac{1}{-\gamma + \sqrt{\gamma^2 + \frac{3}{4}(q^2 - k^2)}}\\
D_s(q;k)&=&\frac{1}{-\gamma_s + \sqrt{\gamma^2 + \frac{3}{4}(q^2 - k^2)}}.
\end{eqnarray}

To obtain Eqs.~(\ref{eq:4.2})--(\ref{eq:4.4}) is now very simple. 
We just define:
\begin{equation}
{\bf X}=-m \left(\begin{array}{cc}
               t_{tt}  & -t_{ts} \\
	       -t_{st} & t_{ss}
		\end{array} \right)
\end{equation}
Using Eq.~(\ref{eq:A8}) to determine the relationship to the doublet
phase shifts and $K$ as in Eq.~(\ref{eq:2.18}) we find that the
t-matrix of Bedaque et al. should obey:
\begin{equation}
t_{tt}(k,k)=\frac{3}{8 \gamma} \frac{1}{k \cot \delta - ik},
\end{equation}
in agreement with Eqs.~(12) and (13) of Ref.~\cite{Bd02}.

\end{document}